\documentclass[
journal=jctcce,
manuscript=article,
layout=onecolumn
]{achemso}

\usepackage[utf8]{inputenc}
\usepackage{amsmath,amsfonts,amssymb}
\usepackage{bm}
\usepackage{xcolor}
\usepackage{float}
\usepackage{graphicx} \graphicspath{{Images/}}
\usepackage{hyperref}
\usepackage[capitalise]{cleveref}
\usepackage{booktabs}
\usepackage[percent]{overpic}
\usepackage{ulem}
\usepackage{empheq} 

\usepackage{framed}
\colorlet{shadecolor}{gray!25} 

\usepackage{nameref}

\usepackage{bbold} 
\usepackage{braket}

\DeclareMathOperator{\Trace}{Tr}

\title{Light-matter hybrid-orbital-based first-principles methods: the influence of the polariton statistics}

\author{Florian Buchholz}
\email{florian.buchholz@mpsd.mpg.de}

\author{Iris Theophilou}
\email{iris.theophilou@mpsd.mpg.de}
\affiliation{Theory Department, Max Planck Institute for the Structure and Dynamics of Matter - Luruper Chaussee 149, 22761 Hamburg, Germany}
\author{Klaas J. H. Giesbertz}
\email{k.j.h.giesbertz@vu.nl}
\affiliation{Department of Theoretical Chemistry and Amsterdam Center for Multiscale Modeling, Faculty of Science, Vrije Universiteit Amsterdam, De Boelelaan 1083, 1081 HV Amsterdam, The Netherlands}
\author{Michael Ruggenthaler}
\email{michael.ruggenthaler@mpsd.mpg.de}
\affiliation{Theory Department, Max Planck Institute for the Structure and Dynamics of Matter - Luruper Chaussee 149, 22761 Hamburg, Germany}
\author{Angel Rubio}
\email{angel.rubio@mpsd.mpg.de}
\affiliation{Theory Department, Max Planck Institute for the Structure and Dynamics of Matter - Luruper Chaussee 149, 22761 Hamburg, Germany}
\altaffiliation{Center for Computational Quantum Physics (CCQ), Flatiron Institute, 162 Fifth Avenue, New York NY 10010, USA}

\keywords{Polaritonic Chemistry, Cavity Quantum Electrodynamics, Electronic Structure Theory, Hartree-Fock, Quantum Optics, Strong Coupling}

\date{\today}

\begin{document}

\newcommand {\td} {\mathrm{d}}
\newcommand {\br} {\mathbf{r}}
\newcommand {\bz} {\mathbf{z}}
\newcommand {\bx} {\mathbf{x}}
\newcommand {\by} {\mathbf{y}}
\newcommand {\blambda} {\bm{\lambda}}
\newcommand{\bq}{\mathbf{q}}

\normalem


\begin{abstract}
A detailed understanding of strong matter-photon interactions requires first-principle methods that can solve the fundamental Pauli-Fierz Hamiltonian of non-relativistic quantum electrodynamics efficiently. A possible way to extend well-established electronic-structure methods to this situation is to embed the Pauli-Fierz Hamiltonian in a higher-dimensional light-matter hybrid auxiliary configuration space. In this work we show the importance of the resulting hybrid Fermi-Bose statistics of the polaritons, which are the new fundamental particles of the ``photon-dressed'' Pauli-Fierz Hamiltonian for systems in cavities. We show that violations of these statistics can lead to unphysical results. We present an efficient way to ensure the proper symmetry of the underlying wave functions by enforcing representability conditions on the dressed one-body reduced density matrix. We further present a general prescription how to extend a given first-principles approach to polaritons and as an example introduce polaritonic Hartree-Fock theory. While being a single-reference method in polariton space, polaritonic Hartree-Fock is a multi-reference method in the electronic space, i.e.\ it describes electronic correlations. We also discuss possible applications to polaritonic QEDFT. We apply this theory to a lattice model and find that the more delocalized the bound-state wave function of the particles is, the stronger it reacts to photons. The main reason is that within a small energy range many states with different electronic configurations are available as opposed to a strongly bound (and hence energetically separated) ground-state wave function. This indicates that under certain conditions coupling to the quantum vacuum of a cavity can indeed modify ground state properties.
\end{abstract}

A plethora of experiments of atoms, molecules and solids embedded in quantum cavities~\cite{Lidzey1998, Chikkaraddy2016,Surface_plasmons, trion-polariton, Kena-Cohen2010, Hutchison2012, manipulation_excited_states, Coles2014transfer, Bayer2017} that were performed in the last two decades, have demonstrated the possibility to change the properties of matter by coupling it strongly to the modes of an optical cavity. In the strong-coupling regime,\footnote{For a definition and detailed discussions of light-matter coupling regimes, see Refs.~\citenum{Kockum2018,Flick2018b, T_rm__2014}} matter degrees of freedom strongly mix with a few effective photon modes such that hybrid light-matter states, called polaritons, emerge. The combined light-matter system can exhibit significantly different properties than the separate subsystems even at ambient conditions, which suggests various interesting applications in chemistry and material science~\cite{Lidzey1998, Chikkaraddy2016,Surface_plasmons}. Examples include the possibility of building polariton lasers~\cite{Kena-Cohen2010}, the modification of chemical landscapes~\cite{Hutchison2012},  selective manipulation of electronic excitations~\cite{manipulation_excited_states}, the control of long-range energy transfer between different matter systems~\cite{Coles2014transfer} or the emergence of new states of matter\cite{Ruggenthaler2018,Kiffner2019,Ashida2020}.
 
The variety of this (by far not exhaustive) list of effects caused by the emergence of polaritons reveals the complexity that arises when light and matter mix strongly. The theoretical description of these effects is far from trivial. Currently, many fundamental questions in the field of polaritonic chemistry are yet to be answered. Among others~\cite{George2016,Herrera2016} are the questions whether collective strong-coupling can modify chemistry without influencing the local electronic structure~\cite{Feist2015,Martinez-Martinez2018} and whether strong coupling to \emph{only} the vacuum field can modify chemical reactions~\cite{Thomas2016}.

Letting aside the cavity setting for a moment, in modern chemistry many fundamental questions about the behavior of matter have been answered by \emph{first-principle} calculations. These well-established methods routinely predict properties of matter, which might interact with ``classical'' time and spatially dependent electromagnetic fields. For example, in order to answer whether a specific chemical reaction happens or not, one calculates the potential-energy surfaces, which allows for the estimation of activation barriers. However, such complex first-principle methods (like Density Functional Theory (DFT)~\cite{Hohenberg1964} or computational quantum-chemistry methods~\cite{Jensen}) are usually geared towards the precise quantum-mechanical description of the electrons (treating ions classically). 
Nevertheless, many phenomena require a quantum mechanical description of the ion dynamics. In a similar way, materials in cavities call for a quantum treatment of the electromagnetic field\cite{Flick2017,Flick2015,Ruggenthaler2014,Galego2019}.

\begin{figure}
	\centering
	\includegraphics[width=0.7\columnwidth]{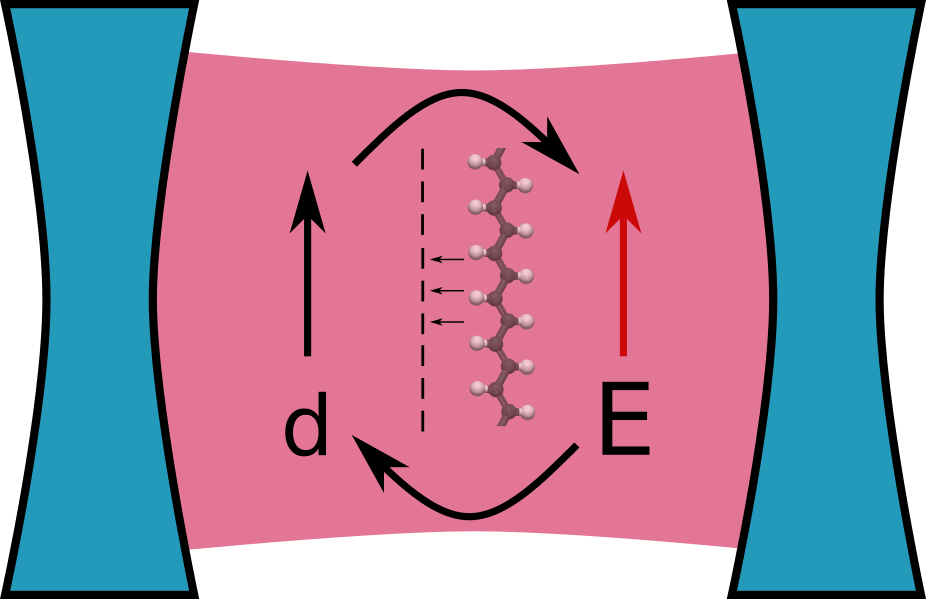}
	\caption{Sketch of our cavity setup. In our explicit example (see Sec.~\ref{sec:study_case}) the electrons of our matter subsystem are allowed to move parallel to the electric field $\mathbf{E}$ of the cavity mode yet restricted in the perpendicular directions, i.e.\ we consider a one-dimensional discretrized matter subsystem. Since the extension in the perpendicular directions is small compared to the wave length of the dominant cavity mode, the coupling is mediated via the total dipole $\mathbf{d}$ of the electrons. If the mode volume (distance between the mirrors) is small or the number of particles increased, new hybrid light-matter quasi-particles, i.e. polaritons, emerge.}
	\label{fig:cavity_sketch}
\end{figure}

In the cavity setting (see Fig.~\ref{fig:cavity_sketch} for a sketch), where besides the electrons and nuclei also the photons play a decisive role, first-principle calculations for only one species of particles are already routinely employed. Although only to provide parameters (such as dipole moments and excitation energies) which serve as essential input in cavity Quantum Electrodynamics (QED) models such as the Rabi and the Jaynes-Cummings model.\footnote{For an overview of such model Hamiltonians, the interested reader is referred to Ref.~\cite{Kockum2018} Box 1.} Photon properties and absorption spectra can be described well within these models in combination with, e.g. input-output theories~\cite{Input-output_theory, Reitz2019}. However, when it comes to polaritonic chemistry, first-principle theories for \emph{light and matter} can provide different (complementary) information~\cite{Ruggenthaler2018}, such as modified Cavity Born--Oppenheimer surfaces~\cite{Flick2017} or the calculation of the most important states in a chemical reaction. Thus, the profit of extending first-principle methods to the cavity setting is twofold: on the one hand they can describe coupled systems parameter-free, i.e.\ less biased, and on the other hand they can provide parameters for cavity QED models~\cite{Wang2020} or even motivate new purpose-build models.
This dual role of first-principle theories is well established, for example, in the context of solid-state physics. In this case, first-principle methods are used either for a full description of the solid or to provide parameters for, e.g.\ the Hubbard models for more complex (strongly-correlated) solids.

The standard electronic-structure methods are geared towards a specific type of particle, i.e., electron or nucleus/ion. And depending on the situation it is usually one of these two particle species that dominates a physical or chemical property. In the case of strong light-matter coupling, where hybrid light-matter states can change these properties, also at least a second particle type, the photon, becomes important as well. Thus one needs to develop specific first-principle methods that can treat several quantized particle species at the same time, such as the Kohn-Sham approach to Quantum-Electrodynamical Density Functional Theory (QEDFT)~\cite{ruggenthaler2011, Tokatly2013,Ruggenthaler2014,jestaedt2019}. 
However, it has been recently shown that first-principles methods, such as QEDFT, can be based on the emerging hybrid particle, the polariton, directly. 
For that the usual dipole-coupled Hamiltonian, i.e. the Pauli-Fierz Hamiltonian in the long-wavelength approximation\cite{Spohn2004}, is embedded into a higher-dimensional polaritonic Hilbert space in an exact way~\cite{Nielsen2018,Buchholz2019}. 
The resulting polaritonic or ``dressed'' Hamiltonian has the same \emph{structure} as the usual matter-only quantum-mechanical Hamiltonian, i.e., one-body kinetic and external potential part and a two-body interaction term with respect to the (now higher dimensional) polaritonic coordinates. This allows one to employ well-established many-body methods, using polaritonic orbitals. This has been demonstrated for time-independent and time-dependent QEDFT~\cite{Nielsen2018} as well as for Hartree-Fock (HF) and Reduced-Density-Matrix Theory (RDMFT)~\cite{Buchholz2019}. Such an approach has several key advantages: One does not need to invent new first-principle methods for the QED setting from scratch,
but can use the machinery of the well-established electronic-structure methods. Further, one works directly in terms of the fundamental quasi-particle that determines the properties of the coupled system. And finally, simple wave functions in terms of polaritonic orbitals correspond to correlated (multi-determinant) wave functions in physical space. 

The above presented approach is exact, since it maps the electron-photon (and more generally the electron-ion-photon) configuration space in a simple and explicit manner to a higher-dimensional polaritonic configuration space. Yet the resulting polaritonic wave functions have now hybrid Fermi-Bose statistics. This has interesting consequences for the first-principle approaches, which are commonly based on Slater determinants of single-particle orbitals. Yet a Slater determinant enforces purely fermionic symmetry. Thus, for the polaritonic extension of first-principle methods we need a construction that enforces this new type of combined fermion-boson symmetry in an easy and efficient manner. While in previous works this has only been approximately enforced and still allowed good results under specific conditions and for model systems~\cite{Nielsen2018,Buchholz2019}, it is to be expected that in general an error in the quasi-particle's statistics can lead to unphysical results. 

In this work we show that indeed approximating the new statistics of the polariton wave function can result in unphysical predictions. Specifically, ignoring the hybrid Fermi-Bose character leads to violations of the Pauli principle, i.e., several electrons can occupy the same quantum state. By then analyzing in detail the hybrid character of the polaritonic wave functions we find necessary conditions that any physical polaritonic wave function needs to obey. In the case of ground-state wave functions, these conditions are sufficient to guarantee the physicality of the wave function. We here show how these conditions can be enforced on the level of reduced density matrices for common first-principle methods with the help of exact inequality constraints on the polaritonic Lagrangian. This result makes such methods a straightforward and reliable tool for the prediction of changes due to strong light-matter interactions. 
As an explicit example we consider how HF theory can be generalized to polaritonic problems and how the inequality constraints lead to new terms in the HF equations that enforce the combined fermion-boson character. We furthermore present numerical results for simple model systems that are, however, already challenging for a straightforward exact numerical calculation. We observe that how multi-electron systems react to strong coupling depends on structural details of the uncoupled system. Most importantly, we find that the more spatially extended the electronic wave function is, the stronger the electrons react to the photons, and the more different the coupled light-matter ground state is when compared to the uncoupled one. This goes together with an increase of the electronic correlations captured by polaritonic HF theory, which is multi-determinantal in the electronic subsystem. These results suggest that coupling to only the vacuum of a cavity can have a strong impact, specifically when the uncoupled electronic wave function is spatially delocalized. This also indicates that when a collective ensemble of emitters is treated from first-principles, there could be strong local modifications, since the collective wave function of all these particles would be highly extended.   

The article is structured as follows. After briefly introducing the standard Hamiltonian of coupled electron-photon systems in Sec.~\ref{sec:physical_setting}, we discuss in detail the polariton statistics of the dressed system and elucidate how and why unphysical solutions may occur without the correct statistics in Sec.~\ref{sec:dressed_orbitals}. We provide instructions on how to generalize a given electronic-structure theory to polaritons with the correct statistics in Sec.~\ref{sec:est}. We conclude this section by explicitly applying our instructions to one of the most ubiquitous approaches, namely HF theory. We then present corresponding numerical results in Sec.~\ref{sec:study_case} and discuss physical implications. We conclude the article with a summary and outlook in Sec.~\ref{sec:conclusion}.

\section{Electrons dipole-coupled to cavity modes}
\label{sec:physical_setting}

As indicated in Fig.~\ref{fig:cavity_sketch}, we consider an atomic or molecular system inside a cavity or, more generally, a nanophotonic environment. Since we assume that the extension of the matter subsystem is small compared to the wavelength of the cavity (perpendicular to the polarization direction of the dominating modes), we can simplify the full minimal-coupling Hamiltonian by adopting the dipole approximation~\cite{craig1998, Spohn2004, jestaedt2019}. If we furthermore assume the nuclei/ions clamped, i.e.\ we work in the Born--Oppenheimer approximation, the problem reduces to $N$ electrons coupled to the cavity modes. While in principle we could also include the nuclei/ions in our description, since the structure of the Hamiltonian (which is the crucial ingredient for the dressed formulation, see Sec.~\ref{sec:dressed_orbitals}) would not change~\cite{schafer2018ab},
we refrain from this more complex situation and concentrate on the electronic structure. In length form and atomic units the Hamiltonian in dipole-approximation for a molecular system in a quantum cavity then reads~\cite{craig1998, Tokatly2013}
\begin{align}\label{eq:Hamiltonian}
    \hat{H} = {}&\underbrace{\sum_{k=1}^N \left[ -\tfrac{1}{2} \nabla_{\br_k}^2 + v(\br_{k}) \right] + \tfrac{1}{2}\sum_{k \neq l}^N w(\br_k,\br_l)}_{\hat{H}_m= \hat{T}[t]+ \hat{V}[v]+ \hat{W}[w]}+\underbrace{\sum_{\alpha=1}^M\left(- \tfrac{1}{2} \tfrac{\partial^2}{\partial p_{\alpha}^2}+ \tfrac{\omega_{\alpha}^2}{2}p_{\alpha}^2\right)}_{\hat{H}_{ph}} \nonumber\\
    &{}+ \underbrace{\sum_{\alpha=1}^M-\omega_{\alpha} p_{\alpha}\blambda_{\alpha} \cdot \hat{\mathbf{D}}}_{\hat{H}_I}
    + \underbrace{\sum_{\alpha=1}^M \tfrac{1}{2}\left(\blambda_{\alpha} \cdot \hat{\mathbf{D}}\right)^2}_{\hat{H}_d}.
\end{align}
The first three terms constitute the usual matter Hamiltonian of quantum mechanics, with the kinetic and external one-body parts, $\hat{T}[t]$ and $\hat{V}[v]$, respectively, and the two-body interaction term $\hat{W}[w]$. Here the kinetic term is the usual Laplacian $t(\br)=-\tfrac{1}{2} \nabla_{\br}^2$, the external potential $v(\br)$ is due to the attractive nuclei/ions and $w(\br,\br')$ is the electron-electron repulsion. Usually this is just taken as the free-space Coulomb interaction $w(\br,\br')=1/|\br-\br'|$, but in a cavity the interaction can be modified~\cite{Power1982}. Since we do not rely on the specific form of the electron-electron repulsion we can, for instance, also easily model the changed interactions due to a plasmonic environment~\cite{Shahbazyan2013}.  
The fourth term $\hat{H}_{ph}$ is the free field-energy of $M$ effective modes of the cavity. The effective modes are characterized by their displacement coordinate $p_{\alpha}$, frequency $\omega_{\alpha}$ and polarization vectors $\blambda_{\alpha}$. The latter include already the effective coupling strength $g_{\alpha}=|\blambda_{\alpha}| \sqrt{\frac{\omega_{\alpha}}{2}}\propto 1/\sqrt{V}$~\cite{Kockum2018, Ruggenthaler2018} that is proportional to the inverse square-root of the cavity mode volume V. In the dipole approximation, the coupling between light and matter is described by the bilinear term $\hat{H}_{I}$ together with the dipole self-energy term $\hat{H}_d$. Here the dipole operator is defined by $\hat{\mathbf{D}}=\sum_{k=1}^N \br_k$. We note that the dipole self-energy term $\hat{H}_d$ is of utmost importance, especially if equilibrium properties are to be considered (as is the case in this work). That is, because this term is responsible for the stability of matter coupled to photon modes~\cite{Rokaj2018} and also guarantees many further fundamental properties of the coupled light-matter system~\cite{Schafer2019}. As a consequence, only by including this (in cavity QED models often discarded) term we can have a well-defined ground state wave function in the basis-set limit.

The corresponding ground-state wave function of Eq.~\eqref{eq:Hamiltonian} depends then on $4N+M$ coordinates 
\begin{align}
\label{eq:wavefunction}
\Psi(\br_1\sigma_1,\dotsc,\br_N\sigma_N;p_1,\dotsc,p_M),
\end{align} 
where $\sigma_k$ are the electronic spin degrees of freedom. The wave function $\Psi$ is anti-symmetric with respect to the exchange of any two electron coordinates $\br_j\sigma_j\leftrightarrow\br_k\sigma_k$, and also depends on $M$ photon-mode displacement coordinates $p_{\alpha}$. The anti-symmetry in $\br \sigma$ enforces the Fermi statistics and thus the Pauli exclusion principle, i.e.\ two electrons cannot occupy the same quantum state. It is important to note that there is no fundamental exchange symmetry between different displacement coordinates. This is due to the fact that the space of each mode is a single-state bosonic Fock space, i.e.\ it counts how many excitations are in that mode, and hence the bosonic symmetry is in the associated creation and annihilation operators (see, e.g., Sec.~2 of~\citet{Buchholz2019}).

Finally, a comment on the basic physical entities in length gauge is in order. Since the length gauge mixes matter and photon degrees of freedom, the displacement coordinates $p_{\alpha}$ do not correspond to the electric field but rather (as the name already indicates) to the auxiliary displacement field~\cite{craig1998, Rokaj2018, Schafer2019}. As a consequence, in contrast to the unitarily equivalent velocity form, $\hat{H}_{ph}$ is not directly proportional to the number of photons but contains matter contributions. Indeed, in length gauge the photon-number operator for mode $\alpha$ is~\cite{Schafer2019}
\begin{align}
\label{eq:PhotonNumber}
	\hat{N}_{ph,\alpha} =& \frac{1}{\omega_{\alpha}} \left(- \frac{1}{2} \frac{\partial^2}{\partial p_{\alpha}^2}+ \frac{\omega_{\alpha}^2}{2}\left(p_{\alpha} - \frac{\blambda_{\alpha}}{\omega} \cdot \hat{\mathbf{D}} \right)^2\right) -\frac{1}{2},
\end{align}
which contains the interaction terms. This highlights that in length gauge we already implicitly work with light-matter quasi-particles. This is, however, not yet sufficient to see that polaritons emerge as the fundamental quasi-particles of the light-matter theory. How this can be made explicit is discussed in the next section.
\section{Dressed electron-photon system}
\label{sec:dressed_orbitals}

Before we discuss the construction that makes the polariton explicitly the fundamental quasi-particle of the dipole-coupled Hamiltonian of Eq.~(\ref{eq:Hamiltonian}), we first want to make some general comments about polaritons and the multi-reference character of coupled electron-photon systems.

As already pointed out before, the hallmark of strong electron-photon coupling is the emergence of light-matter hybrid states or polaritons. The basic models~\cite{Kockum2018} to describe such hybrid states consider two relevant electronic states, labelled by $\ket{g}$ (``ground'') and $\ket{e}$ (``excited'' state), and two photonic states, the vacuum $\ket{0}$ and one-photon state $\ket{1}$. The resulting polaritons are then denoted as
\begin{align}
\label{eq:Polariton_cQED}
P_{\pm}=\alpha \ket{g}\otimes\ket{1} \pm \beta\ket{e}\otimes\ket{0},
\end{align}
with coefficients $\alpha,\beta$ that depend on details of these models. From a quantum-chemical perspective we therefore see that polaritons correspond to \emph{multi-reference} wave functions, even if we described the electronic states as single Slater determinants.
If we go beyond these most simple of models the description of the coupled light-matter states require even more terms for an accurate description, especially for strongly-coupled ground states~\cite{DeLiberato2017}. Systems that require multi-reference states are well known in electronic-structure theory and their accurate description is among the hardest challenges in the field. An example is bond stretching in the hydrogen molecule, where the wave function has a multi-reference character. This prototypical system is commonly used as a challenging test case for first-principle methods~\cite{Handy_dissociation_H2, local_RDMFT, strictly_correlated_H2, mordovina2019self}. However, the term multi-reference depends crucially on the basic entities, i.e.\ the ``single references'',  that are used to build the ``multi-reference'' state. 
Transferred to the problem of polaritonic physics, we face the fundamental problem that separate matter and photon wave functions~\footnote{We want to note that we do not refer to a specific form of ``photon wave functions'' but rather to any basis that is used to expand the photonic part of the coupled electron-photon Hilbert space as, e.g., the photon number states in Eq.~\eqref{eq:Polariton_cQED}.} (as in the simple example above) can be very inefficient in describing polaritonic states of the coupled system. Instead, we want to describe such systems directly by polariton degrees of freedom that depend on both, photonic \emph{and} electronic variables. Specifically, we want to define a one-particle (orbital) basis of polaritons, on which we want to base our considerations. The corresponding polaritonic wave function will differ (however in an explicit and trivial manner, see Eq.~(\ref{eq:SimplePolariton})) from the physical wave function, which we usually construct from separate electronic and photon wave functions. The main technical advantage of this dressed construction, discussed in the following, is that already a single-reference polariton wave function corresponds to a multi-reference coupled electron-photon wave function and hence polaritonic (or dressed) orbitals are potentially more efficient than using the two separate ones.

\begin{figure}
	\centering
	\includegraphics[width=0.7\columnwidth] {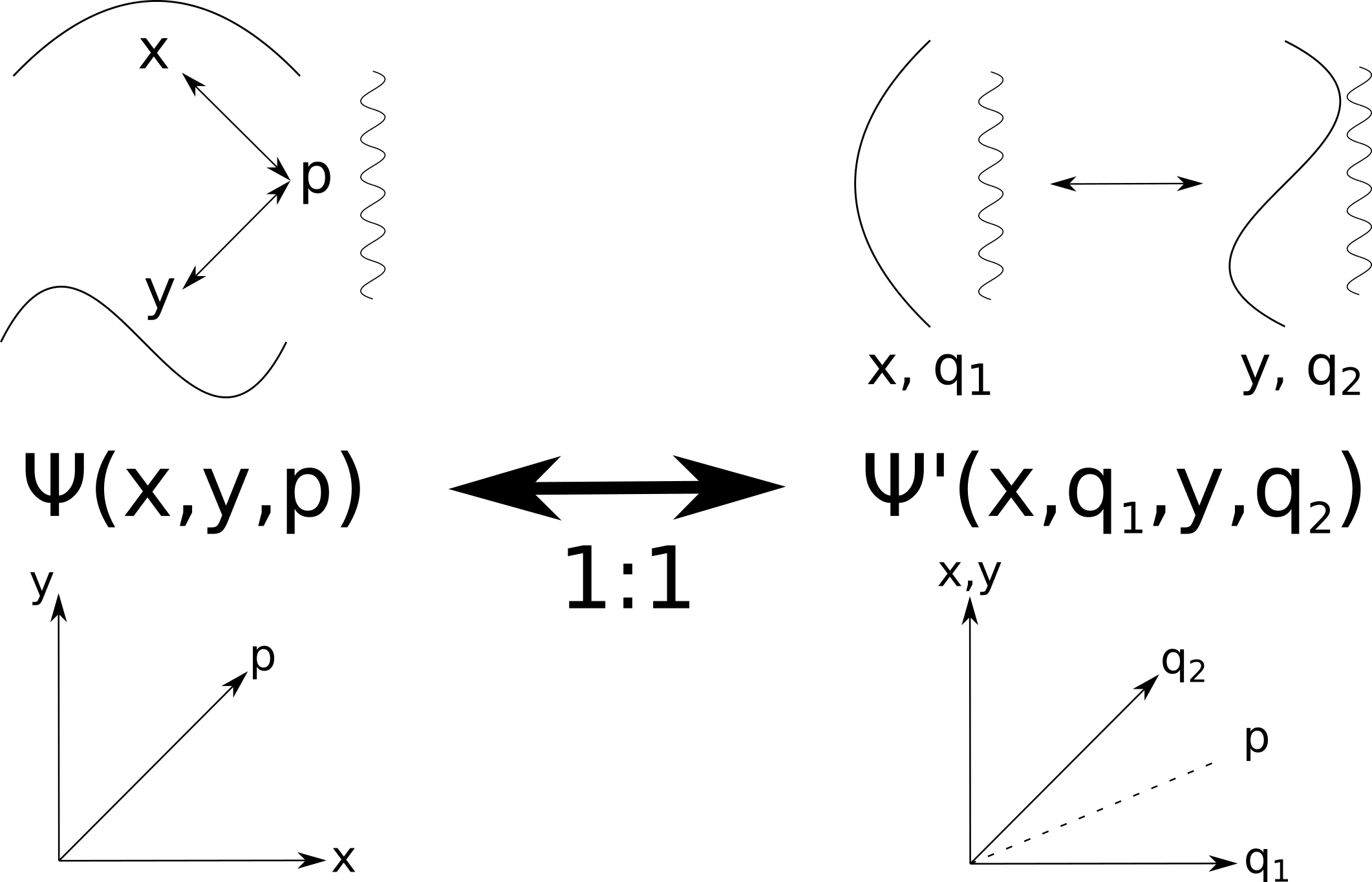}
	\caption{Sketch of the auxiliary construction for an example wave function $\Psi(x,y,p)$ of two one-dimensional electrons $(x,y)$ coupled  to one photon mode with displacement coordinate p. The coupling is indicted by the double arrows $x\leftrightarrow p, y\leftrightarrow p$. The electronic orbital wave functions are symbolized by the ground and first excited state of a box with zero-boundary conditions and the photon mode by a wiggly line. The corresponding dressed wave function $\Psi'(x,q_1,y,q_2)$ has instead two photon-coordinates $q_1,q_2$ that are related to the physical coordinate p by $p=1/\sqrt{2}(q_1+q_2)$. On the wave function level, this connection can be utilized to introduce two polariton orbitals with coordinates $(x,q_1)$ and $(y,q_2)$, respectively, that are interacting. This new interaction is indicated by a double arrow between the two orbitals.}
	\label{fig:dressing_sketch}
\end{figure}
While we focus in this work on the correct hybrid statistics of the polaritons in the dressed construction~\cite{Nielsen2018, Buchholz2019}, we nevertheless want to give a short recapitulation. To turn the coupled electron-photon problem of Sec.~\ref{sec:physical_setting} into an equivalent and exact dressed problem we will follow three steps (for a simplified sketch of the construction, see Fig. \ref{fig:dressing_sketch}): 
\begin{enumerate}

\item For each mode $\alpha=1 ,\dotsc, M$ and all but the first electron $i=2,\dotsc, N$, we introduce extra auxiliary coordinates $p_{\alpha,i}$. This adds $(N-1)M$ extra degrees of freedom to the problem. In this higher-dimensional auxiliary configuration space, we now consider wave functions depending on $4N+NM$ coordinates, i.e.\ 
\begin{align*}
   \Psi' (\br_1 \sigma_1,\dotsc,\br_{N} \sigma_{N}, p_1,\dotsc,p_M, p_{1,2},\dotsc,p_{1,N},\dotsc,p_{M,2}, \dotsc , p_{M,N}).
\end{align*}
Here and in the following, we will denote all quantities in the auxiliary configuration space with a prime.

\item We next construct an auxiliary Hamiltonian in the extended configuration space of the form
\begin{align*}
    \hat{H}'=\hat{H}+\sum_{\alpha=1}^{M} \hat{\Pi}_{\alpha}
\end{align*}
where
\begin{align}
\label{eq:lin_op}
\hat{\Pi}_{\alpha}
= \sum_{i=2}^{N}\biggl( -\frac{1}{2} \frac{\partial^2}{\partial p_{\alpha,i}^2} + \frac{\omega_{\alpha}^2}{2}p_{\alpha,i}^2\biggr)
\end{align}
depends only on these new auxiliary coordinates.  This construction guarantees that the auxiliary degrees of freedom do not mix with the physical ones, which will ensure a simple and explicit connection between the physical and auxiliary system. 
\item Finally we perform an orthogonal coordinate transformation of the physical and auxiliary photon coordinates $(p_1,...,p_{M,N}) \rightarrow (q_{1,1},...,q_{M,N})$ such that 
\begin{align}\label{eq:CenterOfMass}
p_{\alpha} = \tfrac{1}{\sqrt{N}} &\left( q_{\alpha,1} + \dotsb + q_{\alpha,N}  \right), \nonumber\\
-\frac{1}{2} \frac{\partial^2}{\partial p_{\alpha}^2} + \frac{\omega_{\alpha}^2}{2} p_{\alpha}^2 + \hat{\Pi}_{\alpha}& = \sum_{i=1}^{N}\biggl(-\frac{1}{2}\frac{\partial^2}{\partial q_{\alpha,i}^2} + \frac{\omega_{\alpha}^2}{2} q_{\alpha,i}^2\biggr).
\end{align}
Note that the second line is automatically satisfied for any orthogonal transformation and the first line defines $p_{\alpha}$ as the ``center-of-mass'' of all the $q_{\alpha,i}$ with uniform relative masses $1/\sqrt{N}$. 
\end{enumerate}
In total, we then find the auxiliary Hamiltonian in the higher-dimensional configuration space given as
\begin{align}
\hat{H}' {}={}&\sum_{k=1}^N \left[ -\tfrac{1}{2} \nabla_{\br_k}^2 + v(\br_{k}) \right] + \tfrac{1}{2}\sum_{k \neq l} w(\br_k,\br_l)  -\sum_{\alpha=1}^M\omega_{\alpha} p_{\alpha}\blambda_{\alpha} \cdot \hat{\mathbf{D}}+ \sum_{\alpha=1}^M \tfrac{1}{2}\left(\blambda_{\alpha} \cdot \hat{\mathbf{D}}\right)^2 \nonumber\\
&+\sum_{\alpha=1}^M\left(- \tfrac{1}{2} \tfrac{\partial^2}{\partial p_{\alpha}^2}+ \tfrac{\omega_{\alpha}^2}{2}p_{\alpha}^2\right) + \sum_{\alpha=1}^M \sum_{i=2}^{N}\left( -\tfrac{1}{2} \tfrac{\partial^2}{\partial p_{\alpha,i}^2} + \tfrac{\omega_{\alpha}^2}{2}p_{\alpha,i}^2\right)
\nonumber\\
{}\overset{\eqref{eq:CenterOfMass}}{=}{}&\sum_{k=1}^{N}  \left\{-\tfrac{1}{2} \nabla_{\br_k}^2 + v(\br_{k}) + \sum_{\alpha=1}^{M}\! \left[ -\tfrac{1}{2}\tfrac{\partial^2}{\partial q_{\alpha,k}^2} 
+ \tfrac{1}{2}\omega_{\alpha}^2q_{\alpha,k}^2  
- \tfrac{\omega_{\alpha}}{\sqrt{N}} q_{\alpha,k}(\blambda_{\alpha}\! \cdot \br_{k})  
+\tfrac{1}{2}(\blambda_{\alpha}\! \cdot \br_{k})^2 \right]  \right\} \nonumber\\
&+ \tfrac{1}{2}\sum_{k\neq l} \left[ w(\br_{k}, \br_l) + \sum_{\alpha=1}^{M} \left(  - \tfrac{\omega_{\alpha}}{\sqrt{N}} q_{\alpha,k} \blambda_{\alpha} \cdot \br_{l} - \tfrac{\omega_{\alpha}}{\sqrt{N}} q_{\alpha,l} \blambda_{\alpha} \cdot \br_{k}  +  \blambda_{\alpha}\cdot \br_{k} \blambda_{\alpha} \cdot \br_{l}  \right)\right], \nonumber
\end{align}
where we inserted the definition of the total dipole operator and reordered the expressions, such that the terms with only one index and the terms with two different indices are grouped together. 
Introducing then a $(3+M)$-dimensional polaritonic vector of space and transformed photon coordinates $\bz = \br \bq$ with $\bq \equiv (q_1,\dotsc,q_{M})$, we can rewrite the above Hamiltonian as
\begin{align}
\label{AuxiliaryHamiltonian2}
\hat{H}' &= \sum_{k=1}^{N}\left[- \tfrac{1}{2} \Delta_k' + v'(\bz_k) \right] + \tfrac{1}{2} \sum_{k \neq l} w'(\bz_k,\bz_l)\\
&= \hat{T}[t'] + \hat{V}[v'] + \hat{W}[w'], \nonumber
\end{align}
where we introduced the dressed one-body terms
\begin{align}\label{eq:dressedkinetic}
 t'(\bz)&=- \tfrac{1}{2} \Delta_k'\equiv -\tfrac{1}{2}\sum_{i=1}^{3}\tfrac{\partial^2}{\partial r_i^2}-\tfrac{1}{2}\sum_{\alpha=1}^{M}\tfrac{\partial^2}{\partial q_{\alpha}^2},\\ 
 v'(\bz)&=v(\br)+ \sum_{\alpha=1}^{M} \left[ \tfrac{1}{2}\omega_{\alpha}^2q_{\alpha}^2  
- \tfrac{\omega_{\alpha}}{\sqrt{N}} q_{\alpha}\blambda_{\alpha}\!\! \cdot\! \br+\tfrac{1}{2}(\blambda_{\alpha}\!\! \cdot\! \br)^2 \right],\label{eq:dressedpotential}
\end{align}
and the dressed two-body interaction term
\begin{align}\label{eq:dressedinteraction}
    w'(\bz,\bz')= w(\br, \br') + \sum_{\alpha=1}^{M} \left[  - \tfrac{\omega_{\alpha}}{\sqrt{N}} q_{\alpha} \blambda_{\alpha} \!\!\cdot\! \br' - \tfrac{\omega_{\alpha}}{\sqrt{N}} q_{\alpha}' \blambda_{\alpha} \!\!\cdot\! \br  +  \blambda_{\alpha}\!\!\cdot\! \br \blambda_{\alpha} \!\!\cdot\! \br'  \right].
\end{align}
We see here that only the conditions~\eqref{eq:CenterOfMass}, but not the details of the coordinate transformation of step 3 are important for our construction~\cite{Buchholz2019}. We want to stress that the crucial part of this coordinate transformation is the replacement of $p_{\alpha}$ in the interaction terms $p_{\alpha}\blambda_{\alpha} \!\cdot\! \hat{\mathbf{D}}$. Instead of $p_{\alpha}$ only, now all $q_{\alpha,i}$ couple to the dipole of the matter system just with a rescaled coupling-strength by the factor $1/\sqrt{N}$ (see Appendix~\ref{app:mapping} for an explicit example of such a coordinate transformation). Further, as will be discussed in more detail in Sec.~\ref{sec:est}, we are in practice not interested in an exact solution of the dressed problem. Since the auxiliary configuration space is much larger than the original configuration space, we made an (beyond simple systems) unfeasible numerical problem even more unfeasible. Yet, the dressed formulation allows for relatively simple approximation schemes, e.g.\ HF theory in terms of a single polaritonic Slater determinant (see Sec.~\ref{sec:est}). 

To conclude this brief summary of the dressed construction, we want to mention that while the matter observables stay unchanged, the photon observables like the photon energy $\hat{H}_{ph}$ or the displacement coordinate of the photon modes $p_{\alpha}$ have different representations in the auxiliary configuration space. We can, however, explicitly derive the new representations as discussed in detail in \citet{Buchholz2019} The most important examples that we will also use in this work are the photon energy 
\begin{align}
\label{eq:PhotonEnergyConnection}
	E_{ph}=\braket{\hat{H}_{ph}}= \braket{\hat{H}'_{ph}} - \sum_{\alpha=1}^M\frac{N-1}{2}\omega_{\alpha},
\end{align}
that enters the total energy $E=\braket{\hat{H}}=\braket{\hat{H}'}- \sum_{\alpha=1}^M\frac{N-1}{2}\omega_{\alpha}$ and the photon number $N_{ph,\alpha}=\braket{\hat{N}_{ph,\alpha}'}-\frac{N-1}{2}$ of mode $\alpha$. 

Let us next discuss the wave function $\Psi'$ in the auxiliary configuration space. The wave function $\Psi$ in the usual configuration space is a (normalized) solution of the (time-independent) Schrödinger equation $E_0 \Psi = \hat{H} \Psi$. Since $\hat{H}'=\hat{H}+\sum_{\alpha=1
}^M\hat{\Pi}_{\alpha}
$ and $\hat{\Pi}_{\alpha}$  acts only  on the auxiliary coordinates, we can simply construct
\begin{align}\label{eq:SimplePolariton}
\Psi'(\br_1 \sigma_1, \dotsc, p_{M, N}) = \Psi(\br_1 \sigma_1, \dotsc, \br_N \sigma_N;p_1,\dotsc,p_M) \chi(p_{1,2},\dotsc, p_{M, N}),
\end{align}
with $\chi$ being the (normalized) ground state of $\sum_{\alpha=1}^M \hat{\Pi}_{\alpha}$, which is a product of individual harmonic-oscillator ground states. Clearly, $\Psi'$ is a normalized solution of the auxiliary Schrödinger equation $E_0' \Psi' = \hat{H}' \Psi'$. In principle any combination of eigenstates of the auxiliary harmonic oscillators would lead to a new eigenfunction for $\hat{H}'$ but since we here focus on the ground state the natural choice is the lowest-energy solution. Rewriting this wave function in the new coordinates and employing the polaritonic coordinates $\bz \sigma \equiv \br \boldsymbol{q} \sigma$, we arrive at
\begin{align}
\label{eq:DressedManyBodyWF}
\Psi'(\br_1 \sigma_1, \dotsc,\br_N \sigma_N, q_{1,1},\dotsc, q_{M, N}) = \Psi'(\bz_1 \sigma_1,\dotsc,\bz_N \sigma_N).
\end{align}
This polaritonic wave function as the ground state of~\eqref{AuxiliaryHamiltonian2} is the reformulation of the original electron-photon problem of~\eqref{eq:Hamiltonian} we were looking for. Since all the new photonic coordinates belong to harmonic oscillator ground states, exchanging $p_{\alpha,i}$ with $p_{\alpha,j}$ does not change the total wave function $\Psi'$ and this property transfers to the exchange of any coordinate $q_{\alpha,i}$ and $q_{\alpha,j}$. Hence we have now a \emph{bosonic} symmetry with respect to the $\boldsymbol{q}$ coordinates. Since the electronic part of the auxiliary system is not affected by the coordinate transformation, the electronic symmetries are the same in the physical and auxiliary system, i.e.\ we have a \emph{fermionic} symmetry with respect to $\br \sigma$. Together these two fundamental symmetries imply that the polaritonic coordinates $\bz \sigma$ have fermionic character. The symmetries of the polaritonic wave function $\Psi'$ can  be summarized as
\begin{subequations}
\label{eq:SymmetryPauliDressedFermion}
\begin{empheq}[box=\fbox]{alignat=3}
	\br_k \sigma_k&\leftrightarrow \br_l \sigma_l \quad&\rightarrow\quad \Psi'&\leftrightarrow-\Psi' \\
	\boldsymbol{q}_k&\leftrightarrow\boldsymbol{q}_l &\rightarrow\quad \Psi'&\leftrightarrow\Psi'
\end{empheq}
\end{subequations}
from which follows
\begin{empheq}[box=\fbox]{align}\label{eq:DressedFermionApproximation}
	\bz_k \sigma_k\leftrightarrow \bz_l \sigma_l \quad&\rightarrow\quad \Psi'\leftrightarrow-\Psi'
\end{empheq}
This means that though  the dressed wave function has fermionic statistics~\eqref{eq:DressedFermionApproximation} in terms of the polaritonic coordinates $\bz \sigma$, due to the \emph{constitutive relations}~\eqref{eq:SymmetryPauliDressedFermion} it actually consists of two types of particles: one with fermionic character and another with bosonic character. Consequently, the polariton wave function $\Psi'$ has a hybrid Fermi-Bose statistics. As consequences of these symmetries we find the Pauli exclusion principle for the electrons, yet for the auxiliary photon coordinates we find that many photonic auxiliary entities can occupy the same quantum state. 

Let us briefly highlight that the constitutive relations and the hybrid fermion-boson symmetry are fundamental to the polaritonic wave function. The fermionic statistics of Eq.~\eqref{eq:DressedFermionApproximation}, which are merely a consequence of the constitutive relations, are in general not sufficient to guarantee a physically reasonable result. Let us illustrate this with a simple two-electron-one-mode example with no electron-electron and electron-photon interactions. The exact physical ground state
\begin{align*}
\Psi(\br_1 \sigma_1,\br_2 \sigma_2,p) = \frac{1}{\sqrt{2}}\left[\vphantom{\sum}\psi_1(\br_1 \sigma_1)\psi_2(\br_2 \sigma_2)- \psi_2(\br_1 \sigma_1)\psi_1(\br_2 \sigma_2)\right] \chi_0(p),
\end{align*}
is just a product of a Slater determinant consisting of the two lowest eigenfunctions $\psi_{1},\psi_2$ of $\hat{T}[t]+\hat{V}[v]$ and $\chi_0(p)=(\frac{\omega}{\pi})^{1/4} e^{-\omega p^2/4}$, which is the ground state of a harmonic oscillator with frequency $\omega$. We obtain the dressed version of $\Psi$ by multiplication with another oscillator ground state $\chi_0(p_2)$ with the same frequency. Performing the coordinate transformation~\eqref{eq:CenterOfMass} in this case simply replaces the coordinates $(p,p_2)$ with $(q_1,q_2)$ since they are orthogonal.\footnote{Specifically, we need to perform the transformation for the term $\chi_0(p)\chi_0(p_2)\propto e^{-\omega p^2/2}e^{-\omega p_2^2/2}$. Since an orthogonal transformation for a set of variables $(x_1,x_2,...)\rightarrow(x_1',x_2',...)$ leaves any expression of the form $\sum_i x_i^2\rightarrow\sum_i x_i'{}^2$ invariant, we have $e^{-\omega p^2/2}e^{-\omega p_2^2/2}\rightarrow e^{-\omega q_1^2/2}e^{-\omega q_2^2/2}$.} Note that for a coupled problem the transformation becomes more involved (see Appendix~\ref{app:mapping} for details). The correct auxiliary ground state reads
\begin{align*}
	\Psi'(\bz_1 \sigma_1,\bz_2 \sigma_2) &=\frac{1}{\sqrt{2}} \left[\vphantom{\sum}\psi_1(\br_1 \sigma_1)\psi_2(\br_2 \sigma_2)- \psi_2(\br_1 \sigma_1)\psi_1(\br_2 \sigma_2)\right] \chi_0(q_1) \chi_0(q_2) \nonumber \\
	&=\frac{1}{\sqrt{2}} \left[\vphantom{\sum}\psi_1(\br_1 \sigma_1)\chi_0(q_1) \psi_2(\br_2 \sigma_2)\chi_0(q_2) - \psi_2(\br_1 \sigma_1)\chi_0(q_1)\psi_1(\br_2 \sigma_2)\chi_0(q_2)\right] \nonumber\\
	&=\frac{1}{\sqrt{2}} [\phi_{10}(\bz_1 \sigma_1)\phi_{20}(\bz_2 \sigma_2)- \phi_{20}(\bz_1 \sigma_1)\phi_{10}(\bz_2 \sigma_2)],
\end{align*}
where in the last line we subsumed the electronic and photonic orbitals with the same coordinate index to a polariton orbital, i.e.\ $\psi_i(\br_j \sigma_j)\chi_n(q_j) \equiv\phi_{in}(\bz_j \sigma_j)$. This wave function is obviously antisymmetric with respect to the exchange of $\bz_1 \sigma_1$ and $\bz_2 \sigma_2$ but it also obeys the constitutive relations that enforce the hybrid symmetry in $\br \sigma$ and $q$, respectively. Now let us consider a possible wave function that \emph{only} obeys the overall fermionic symmetry, e.g.\
\begin{align*}
	\tilde{\Psi}'(\bz_1 \sigma_1,\bz_2 \sigma_2)&= \phi_{10}(\bz_1,\sigma_1)\phi_{11}(\bz_2,\sigma_2)- \phi_{11}(\bz_1,\sigma_1)\phi_{10}(\bz_2,\sigma_2) \\
	&= \psi_1(\br_1 \sigma_1)\psi_1(\br_2 \sigma_2) \left[\vphantom{\sum}\chi_0(q_1) \chi_1(q_2) - \chi_1(q_1)\chi_0(q_2)\right].
\end{align*}
Without enforcing also the hybrid statistics both wave functions would be possible eigenfunctions of the non-interacting and uncoupled auxiliary Hamiltonian $\hat{H}'$. Depending on their energy eigenvalues $E[\Psi']$, a minimization of the dressed Hamiltonian without ensuring the hybrid statistics could determine either of the two wave functions as ground state. Only if $E [\Psi']=\epsilon_1+\epsilon_2+\omega< 2 \epsilon_1+ 2\omega =E [\tilde{\Psi}']$, where $\epsilon_1 $ and $\epsilon_2$ are the eigenenergies corresponding to $\psi_1$ and $\psi_2$, a simple minimization would yield the right symmetry solution. If, however, $\epsilon_2-\epsilon_1 > \omega$, then a minimization of the dressed problem without further symmetry restrictions would lead to the state $\tilde{\Psi}'$ that violates the Pauli principle. For a coupled problem, i.e.\ $\lambda >0$, both cases cannot be separated so easily, but the problem remains in principle the same (see Sec.~\ref{sec:study_case} for details). Thus, we either have to make sure that $\omega$ is large compared to the electronic excitations, such that the unrestricted minimization with \emph{only} fermionic symmetry in $\bz \sigma$ picks the right wave function~\cite{Buchholz2019}, or we have to enforce the hybrid statistics.
Indeed, if we enforce the constitutive relations also on $\tilde{\Psi}'$ by adding two extra terms, we see that
\begin{align*}
\tilde{\Psi}'_{hybrid} (\bz_1 \sigma_1,\bz_2 \sigma_2) {}={}& \overbrace{\phi_{10}(\br_1 q_1 \sigma_1)\phi_{11}(\br_2 q_2 \sigma_2)- \phi_{11}(\br_1 q_1 \sigma_1)\phi_{10}(\br_2 q_2 \sigma_2)}^{	\tilde{\Psi}'} \\
&+ \phi_{10}(\br_1 q_2 \sigma_1)\phi_{11}(\br_2 q_1 \sigma_2)- \phi_{11}(\br_1 q_2 \sigma_1)\phi_{10}(\br_2 q_1 \sigma_2)\\
{}={}& \psi_1(\br_1 \sigma_1)\psi_1(\br_2 \sigma_2) \left[\vphantom{\sum}\chi_0(q_1) \chi_1(q_2) - \chi_1(q_1)\chi_0(q_2)\right] \\
&+ \psi_1(\br_1 \sigma_1)\psi_1(\br_2 \sigma_2) \left[\vphantom{\sum}\chi_0(q_2) \chi_1(q_1) - \chi_1(q_2)\chi_0(q_1)\right] \\
{}={}& 0.
\end{align*}
Thus we obtain the desired result of ruling out the solution that violates the Pauli principle. It can be shown~\cite{Buchholz2019} that enforcing the conditions of Eq.~\eqref{eq:SymmetryPauliDressedFermion} is sufficient for the dressed system to attain as ground state $\Psi'=\Psi \chi$, where $\Psi$ is the ground state of the original Hamiltonian of Eq.~\eqref{eq:Hamiltonian} and $\chi$ the product of ground-state harmonic oscillators. For excited states, the constitutive relations are necessary but not sufficient to single out the eigenfunctions of $\hat{H}'$ that correspond to the original Hamiltonian in terms of simple products~\cite{Nielsen2018}. 

So, how can we enforce the right symmetry and with this the right hybrid statistics in practice? We would like to use the polaritonic orbitals $\phi_i(\bz \sigma)$ as our basic physical entity and this makes the constitutive relations quite special. This is due to the fact that they concern always just a \emph{subset of the coordinates} of one orbital. Contrary to the usual approach in many-body physics, where a single-orbital basis is chosen and the fermionic/bosonic symmetry is enforced by building Slater determinants/permanents. One could try to transfer this approach to the polariton case and construct a many-body basis with elements that satisfy Eq.~\eqref{eq:SymmetryPauliDressedFermion} and thus are Slater determinants not only with respect to the polaritonic, but also with respect to the electronic coordinates (or equivalently permanents with respect to the photonic coordinates). This would lead to basis elements with ``mixed-index'' orbitals, i.e.\ orbitals $\phi(\br_i,q_j,\sigma_i)$ that depend on coordinates with \emph{different} indices $i\neq j$. If we want to calculate the expectation value of an observable and integrate over all coordinates, we see that coordinate $i$ and $j$ are coupled alone due to the ``mixed-index'' orbital. This has severe consequences. For example, we cannot make use of the orthonormality relation to set certain terms to zero when we calculate matrix elements and one-body terms become ``two-body like.'' The number of such anomalous terms grows \emph{factorially} with the particle number, which makes such a type of ansatz infeasible in practice. For details the reader is referred to Appendix~\ref{app:symmetry}.

As an alternative to looking at the polaritonic wave function directly, the physical conditions \eqref{eq:SymmetryPauliDressedFermion} are visible in the dressed one-body reduced density matrix (1RDM)~\cite{Buchholz2019}, which is given explicitly by
\begin{align}
\gamma[{\Psi'}](\bz \sigma,\bz'\sigma')= \sum_{\sigma_2,\dotsc,\sigma_N}\int\td^{3(N-1)}\bz \Psi'{}^*(\bz' \sigma',\dotsc,\bz_N \sigma_N) \Psi'(\bz \sigma,\dotsc,\bz_N \sigma_N).  
\end{align} 
The Fermi statistics of the wave function $\Psi'$ with respect to the polaritonic coordinates $\bz\sigma$~\eqref{eq:DressedFermionApproximation} is also apparent in $\gamma[\Psi']$ in the form of so-called $N$-representability conditions\cite{Coleman1963}.
By using the \emph{natural orbitals} $\phi_i$ and the \emph{natural occupation numbers} $n_i$, which are defined by the eigenvalue equation $n_i \,\phi_i = \hat{\gamma} \,\psi_{i}$, we represent $\gamma$ in its diagonal form
\begin{align}
\label{eq:Dressed1RDM_diagonal_form}
\gamma[{\Psi'}](\bz \sigma, \bz' \sigma') =\sum_{i=1}^{\infty} n_i \, \phi_i^{*}(\bz' \sigma') \phi_i(\bz \sigma).
\end{align}
The fermionic $N$-representability conditions become especially simple in this representation and are given by
\begin{empheq}[box=\fbox]{align}
\label{eq:NrepDressed1RDM}
\begin{split}
&0 \leq \,n_i \leq 1, \quad \forall i \\
&\textstyle\sum_i n_i = N.
\end{split}
\end{empheq}
In general, it can be proven\cite{Coleman1963} that \emph{any} matrix that fulfils the conditions~\eqref{eq:NrepDressed1RDM} is connected to an ensemble of $N$-body states that are fermionic with respect to the exchange of their coordinates. 

From the dressed 1RDM we can define the electronic 1RDM
\begin{align}\label{eq:GammaElectronic}
\gamma_e[{\Psi'}](\br \sigma, \br' \sigma') = \int \td^M \bq \;\gamma[{\Psi'}](\br \bq \sigma,\br' \bq \sigma')
\end{align}
and the auxiliary photonic 1RDM
\begin{align}
\gamma_p[{\Psi'}](\bq, \bq') = \sum_{\sigma}\int \td^3 \br \, \gamma[{\Psi'}](\br \bq \sigma,\br \bq' \sigma).
\end{align}
Again, we can define the according natural orbitals $\psi_i^{e/p}$ and the natural occupation numbers $n_i^{e/p}$ by the eigenvalue equations $n_i \psi_i^{e/p} = \hat{\gamma}_{e/p} \psi_{i}^{e/p}$ and go into their diagonal representations
\begin{align}
\label{eq:gamma_electronic_eigenrep}
\gamma_e[{\Psi'}](\br \sigma, \br' \sigma') =\sum_{i=1}^{\infty} n_i^{e} \, \psi_i^{e}{}^*(\br' \sigma') \psi_i^{e}(\br \sigma)
\end{align}
and
\begin{align}
\gamma_p[{\Psi'}](\bq, \bq') =\sum_{i=1}^{\infty} n_i^{p} \, \psi_i^{p}{}^*(\bq') \psi_i^{p}(\bq).
\end{align}
The Fermi statistics with regard to only the electronic coordinates $\br\sigma$ thus becomes apparent by considering the electronic natural occupation numbers $n^e_i$
\begin{subequations}
	\label{eq:Nrepresentability}
\begin{empheq}[box=\fbox]{align}
	\label{eq:Nrepresentability1}
	&n_i^{e} \geq \,0, \quad \forall i \\
	\label{eq:Nrepresentability2}
	&n_i^{e} \leq 1, \quad \forall i \\
	\label{eq:Nrepresentability3}
	&\textstyle\sum_i n_i^{e} = N,
\end{empheq}
\end{subequations}
where we split the conditions in three parts for later convenience. The equivalent bosonic symmetry of the auxiliary photonic coordinates leads instead to the conditions 
\begin{subequations}
	\label{eq:NrepresentabilityEigenrepresentationPhotons}
\begin{empheq}[box=\fbox]{align}
\label{eq:NrepresentabilityEigenrepresentationPhotons1}
&0 \leq \,n_i^{p} , \quad \forall i \\
\label{eq:NrepresentabilityEigenrepresentationPhotons2}
&\textstyle\sum_i n_i^{p} = N.
\end{empheq}
\end{subequations}
Note that the normalization of $\gamma_{e/b}$ to the electron number $N$ is a direct consequence of the auxiliary construction that considers exactly $N$ polaritons for a system with $N$ electrons. This becomes explicitly visible in the fact that the normalization of $\gamma$ by definition transfers to $\gamma_{e/b}$, since $N=\sum_{\sigma}\int\td\bz\gamma(\bz\sigma,\bz\sigma)=\sum_{\sigma}\int\td\br\gamma_e(\br\sigma,\br\sigma)= \int\td\bq\gamma_p(\bq,\bq)$. Additionally, the lower bounds of $\gamma_{e/b}$, c.f. Eqs.~\eqref{eq:Nrepresentability1} and~\eqref{eq:NrepresentabilityEigenrepresentationPhotons1}, transfer from $\gamma$, because the partial trace operation is a completely positive map~\cite{Pillis1967}. We can conclude that if~\eqref{eq:NrepDressed1RDM} is enforced, only the upper bound of the electronic 1RDM, c.f. Eq.~\eqref{eq:Nrepresentability2} provides a non-trivial additional constrained.

This now shows explicitly also for an interacting wave function that at most one electron can occupy a specific quantum state, while many auxiliary photon quantities can occupy a single quantum state.  Further, the dressed 1RDM $\gamma[\Psi']$ itself has only natural occupation numbers between zero and one and is therefore fermionic, yet it contains a fermionic and a bosonic subsystem. It is important to note that the original wave function $\Psi$ did not have this simple hybrid statistics but only fermionic symmetry, since the physical $p_\alpha$ did not follow any specific statistics. Further, that we genuinely have formulated the coupled electron-photon problem in terms of hybrid quasi-particles becomes most evident by actually using single-particle (polariton) orbitals $\phi_i(\bz \sigma)$ to expand the dressed 1RDM of $\Psi'$. Why we consider the 1RDM and how we can do this efficiently will be discussed in the following.

Let us illustrate this with the example from before. The correct auxiliary ground state $\Psi'$ satisfies the conditions of Eq.~\eqref{eq:Nrepresentability}, since
\begin{align*}
 \gamma_e[{\Psi'}]=&\sum_{\sigma_2}\int\td\bz_2\td q_1 {\Psi'}^*(\br' q_1 \sigma',\bz_2 \sigma_2) \Psi'(\br q_1 \sigma,\bz_2 \sigma_2) \\
 =& \sum_{i=1}^{2} \psi_i^*(\br'\sigma')\psi_i(\br\sigma).
\end{align*}
The two electronic orbitals are the eigenfunctions (natural orbitals) of $\gamma_e[\Psi']$ with natural occupation numbers $n_1=n_2=1$. If we do the same calculation with $\tilde{\Psi}'$, we get instead
\begin{align*}
	\gamma_e[ {\tilde{\Psi}'}] = 2 \psi_1^*(\br' \sigma')\psi_1(\br \sigma),
\end{align*}
which violates the $N$-representability conditions~\eqref{eq:Nrepresentability} and thus the Pauli principle. For more intricate wave functions, the diagonalization of $\gamma_e[\Psi']$ will not be as trivial as for this simple example, but nevertheless the conditions~\eqref{eq:Nrepresentability} are \emph{sufficient} to ensure the Pauli exclusion principle in the sense that maximally one fermion can occupy a single quantum state, c.f. Eq.~\eqref{eq:Nrepresentability2}. The conditions~\eqref{eq:Nrepresentability} are however not sufficient to guarantee that there is a pure state $\ket{\Psi'}$ which yields this 1RDM. Additional constraints would need to be invoked which rapidly grow in complexity to ensure that the 1RDM is \emph{pure state} $N$-representable\citep{Klyachko2006}. Nevertheless, the conditions~\eqref{eq:Nrepresentability} are sufficient to guarantee that an ensemble (mixed state) $\Gamma = \sum_{j}w_j \ket{\Psi'_j}\bra{\Psi_j'}$ with $\sum_{j}w_j =1$ exists, which yields this 1RDM\citep{Coleman1963}. The 1RDM is then said to be \emph{ensemble} $N$-representable. Thus, though~\eqref{eq:Nrepresentability} does not guarantee that $\Psi'$ satisfies~\eqref{eq:SymmetryPauliDressedFermion}, it does guarantee that it satisfies the Pauli exclusion principle and additionally that an ensemble constructed from $\ket{\Psi'_j}$ which do satisfy~\eqref{eq:SymmetryPauliDressedFermion} exists.

To obtain a computationally tractable procedure, we therefore use the construction presented in the next section to ensure the polariton statistics implied by Eq.~\eqref{eq:SymmetryPauliDressedFermion}, instead of the \emph{factorially} growing number of ``mixed-index'' orbitals. We will consider all fermionic density matrices in the auxiliary configuration space, which we characterize by the conditions of Eq.~\eqref{eq:NrepDressed1RDM} in terms of polaritonic orbitals $\phi_i(\bz \sigma)$. We then constrain this space by enforcing the $N$-representability conditions of Eq.~\eqref{eq:Nrepresentability} for the 1RDM of the electronic subsystem. Since this guarantees that only (ensembles of) fermionic wave functions are allowed, also the minimal energy solution has fermionic symmetry with respect to $\br \sigma$. This together with the $\bz\sigma$ antisymmetry implies that $\mathbf{q}$ has bosonic symmetry. We call this construction the polariton ansatz for strong light-matter interaction. In the next section, we will based on the polariton ansatz provide a detailed prescription to generalize a given electronic-structure theory to treat ground states of coupled electron-photon systems from first principles. 
\section{First-principle theories: from electronic to polaritonic bases}
\label{sec:est}
In this section, we lay out in detail how one can transform a given electronic-structure theory that meets some minimal requirements into its polaritonic version. The goal of such a ``polaritonic-structure theory'' is to find the ground state of the Hamiltonian of Eq.~\eqref{eq:Hamiltonian} by considering the ground state of the auxiliary Hamiltonian of Eq.~\eqref{AuxiliaryHamiltonian2}. We define the according variational principle for the ground-state energy $E_0'$ as
\begin{align}
\label{eq:VariationalPrinciplePauliDressedFermions}
	E_0' = \inf_{\Psi' \,\in\, \mathfrak{P}} \braket{\Psi' |\hat{H}'\Psi'},
\end{align}
where $\mathfrak{P}=\{\Psi' : \Psi' \leftrightarrow \text{\eqref{eq:SymmetryPauliDressedFermion}} \}$ is the set of all normalized many-polariton wave functions that obey the constitutive relations of Eq.~\eqref{eq:SymmetryPauliDressedFermion}. For our purposes, as explained in Sec.~\ref{sec:dressed_orbitals}, we will instead consider the larger set $\mathfrak{M}$ of all (mixed-state) density matrices $\Gamma = \sum_{j}w_j \ket{\Psi'_j}\bra{\Psi_j'}$ with $\sum_{j}w_j =1$, that obey the hybrid Fermi-Bose statistics. The minimal energy also in this more general set corresponds to the pure state of Eq.~\eqref{eq:VariationalPrinciplePauliDressedFermions}, i.e. 
\begin{align}
\label{eq:VariationalPrinciplePauliDressedFermionsGamma}
	E_0' = \inf_{\Gamma \,\in\, \mathfrak{M}} \Trace\{ \Gamma \hat{H} \}.
\end{align}
The main trick now is in how we construct this set. We do so by first considering the yet larger set $\tilde{\mathfrak{M}} = \{\tilde{\Gamma} : \tilde{\Psi'_j} = \sum C_{j,K} \Phi_K \}$, i.e.\ density matrices made of superpositions of Slater determinants $\Phi_K=\text{det}(\phi_{K,1}\cdots\phi_{K,N})/\sqrt{N!}$ of polariton orbitals $\phi_{K,i}$. This guarantees the overall Fermi statistics in terms of the polaritonic coordinates $\bz \sigma$. We then constrain this larger set to
\begin{align}
\label{eq:PolaritonSpace}
	\mathcal{M}=\{\tilde{\Gamma} \in \tilde{\mathfrak{M}} :  n^e_i[{\tilde{\Gamma}}] \leq 1 \},
\end{align}
where $n^e_i[{\tilde{\Gamma}}]$ are the natural occupation numbers, c.f. Eq.~\eqref{eq:gamma_electronic_eigenrep}, of the electronic 1RDM $\gamma_e[\tilde{\Gamma}] = \sum_{j} w_j \gamma_e[\tilde{\Psi'_j}]$ that depend on $\tilde{\Gamma}$. This enforces the fermionic statistics with respect to the electronic coordinates $\br \sigma$. The rest of the $N$-representability conditions (Eqs.~\eqref{eq:Nrepresentability1} and~\eqref{eq:Nrepresentability3}) are satisfied automatically by choosing $\Psi'\in \tilde{\mathcal{P}}$ as fermionic with respect to polariton coordinates and thus the corresponding dressed 1RDM satisfies the $N$-representability conditions (Eq. \eqref{eq:NrepDressed1RDM}). As we explained in Sec. \ref{sec:dressed_orbitals}, this is sufficient to ensure the constitutive relation \eqref{eq:SymmetryPauliDressedFermion}. However, this does not automatically imply that the wave functions $\tilde{\Psi'_{j}}$ that are used to construct the constrained density matrices obey the basic symmetries. Rather, they are also auxiliary quantities and it is only the density matrices that are the physical objects. Due to the ensemble conditions there are, however, wave functions with the exact hybrid symmetries associated. We thus avoid the direct construction of the exponentially growing correlated electron-photon states.\footnote{Strictly speaking, we should construct the ensembles $\Gamma = \sum_jw_j \ket{\Psi'_j}\bra{\Psi_j'}$ which generates $\gamma[\tilde{\Psi}']$ to evaluate the energy as $\Trace\{ \Gamma \hat{H} \}$ to remain variational~\eqref{eq:VariationalPrinciplePauliDressedFermionsGamma}. But instead we evaluate the energy directly from $\tilde{\Psi}'$ which only satisfies~\eqref{eq:Nrepresentability} as $\braket{\tilde{\Psi}' | \hat{H}'\tilde{\Psi}'}$. However, because $\gamma(\bz\sigma,\bz'\sigma')$ is correct, the error is only in the correlation part of the two-body part of the energy. This means that for electronic structure theories only based on the 1RDM, e.g.\ Kohn--Sham DFT, HF and RDMFT, it is exact, since only the 1RDM is relevant in those theories.}

Most importantly, the polariton picture gives any coupled problem of the form of Eq.~\eqref{eq:Hamiltonian} the \emph{same structure} as a purely electronic problem with two-body interactions. Consequently, we can transfer every type of electronic-structure theory to the coupled electron-photon problem, if the theory provides an expression for the 1RDM (since we need the 1RDM to test the N-representability constraints). 
\begin{figure}
	\fbox{
	\begin{tabular}{cccl}
		Hamiltonian & Basis States & & Lagrangian of the EST \\
		\rule{0pt}{15pt}
		$\hat{H}_m = \hat{T}[t] +\hat{V}[v]+\hat{W}[w]$ & 
		$\{\phi_k(\br\sigma)\}$ &$\rightarrow$ 
		& $L_m[\Psi] = E_m^{t,v,w}[\Psi] + \mathcal{C}[\Psi]$ \\
		\rule{0pt}{15pt}
		$\downarrow$ & & & \\
		\rule{0pt}{15pt}
		$\hat{H}=\hat{H}_m + \hat{H}_{ph} + \hat{H}_I + \hat{H}_d$ & 
		$\{\phi_k(\br\sigma),\chi_{\alpha}(p)\}$ &$\rightarrow$ & ``new theory'' \\
		\rule{0pt}{15pt}
		$\downarrow$ & & & \\
		\rule{0pt}{15pt}
		$\hat{H}'= \hat{T}[t'] +\hat{V}[v']+\hat{W}[w']$ &
		$\{\phi_k'(\br\sigma,q)\}$ &$\rightarrow$ & $L'[\Psi'] = E_m^{t',v',w'}[\Psi'] + \mathcal{C}[\Psi'] + \mathcal{G}[\Psi']$
	\end{tabular}
	}
	\caption{Graphical illustration of the polariton construction and its connection to an electronic-structure theory (EST). Here $E_m$ indicates the energy expression of the EST, such as the HF, configuration interaction, or coupled cluster energy functional, and $\mathcal{C}$ indicates the constraints of the EST, such as orthonormality of the orbitals. They are enforced on a (possibly multi-determinantal) wave function $\Psi$ constructed from an electronic single-particle basis $\phi_k$. Further, $\mathcal{G}$ indicates the new constraints that arise due to the hybrid statistics of the polaritons, which are now enforced on a (possibly multi-determinantal) wave function $\Psi'$ of polaritonic single-particle orbitals $\phi'_k$. For the usual coupled electron-photon problem (second line), whose Hamiltonian has a different structure and is build on separate orbitals $\phi_k$ and $\chi_\alpha$. Thus a new (efficient and accurate) approximate energy expression would be needed.}
	\label{fig:polariton_construction}
\end{figure}
The main steps how to do so are depicted in Fig.~\ref{fig:polariton_construction}. We assume that the theory provides us an energy expression $E_m$ with respect to a set of electronic basis states $\{\phi_k\}$. This requirement is met by basically every electronic structure theory, as for instance Kohn--Sham DFT or HF but also coupled cluster, valence bond theory or configuration interaction. Depending on the specific theory, $E_m$ might have quite different forms, but it is always derived from some many-body Hamiltonian $\hat{H}_m=\hat{T}[t]+\hat{V}[v]+\hat{W}[w]$. More specifically, the connection between $\hat{H}_m$ and $E_m$ is given by the particle number $N$ and the integral kernels $(t,v,w)$ of the three energy operators. For the matter Hamiltonian $\hat{H}_m$ of Eq.~\eqref{eq:Hamiltonian} for example, these kernels are given by $t=-\frac{1}{2}\nabla^2_{\br}$, $v=v(\br)$ and $w=w(\br,\br')$. The goal of any electronic structure theory is then to find the minimum of $E_m^{t,v,w}[\Psi]$, where $\Psi$ is a (possibly multi-determinantal) wave function constructed from the orbital set $\{\phi_k\}$. 
Typically, one needs to impose some constraints on the parametrization of the wave function $\Psi$ to make it physical $c_k[\Psi] = 0$, e.g.\ orthonormality of orbitals or the norm of the CI coefficients.
The \emph{generic electronic-structure minimization problem} is formulated as
\begin{alignat}{2}
\label{eq:MinimizationProblemEST}
&\text{minimize }\, &&E_m^{t,v,w}[\Psi] \nonumber \\
&\text{subject to }\quad &&c_k[\Psi] = 0.
\end{alignat}
We can solve Eq.~\eqref{eq:MinimizationProblemEST} by, e.g.\ minimizing the Lagrangian
\begin{align}
	L^{t,v,w}_m[\Psi,\{\epsilon_k\}] = E_m^{t,v,w}[\Psi] + \mathcal{C}[\Psi,\{\epsilon_k\}],
\end{align}
where $\mathcal{C}[\Psi,\{\epsilon_k\}] = \sum_{k}\epsilon_k c_{k}[\Psi]$ is a Lagrange-multiplier term. Instead of minimizing $E_m$ directly, one minimizes $L_m$ with respect to the orbitals and the Lagrange-multipliers $\epsilon_k$. Today, a plethora of standard electronic-structure codes exist that solve \eqref{eq:MinimizationProblemEST} very efficiently for many different theory levels and thus allow for a highly accurate description of electronic structure.


If we consider the coupled electron-photon Hamiltonian of Eq.~\eqref{eq:Hamiltonian} instead of the purely electron Hamiltonian $\hat{H}_m$, we find that we need to build new approximation strategies and implementations to deal with the coupled electron-photon Hamiltonian directly. However, by transforming the problem into its dressed counterpart, i.e.\ we consider~\eqref{AuxiliaryHamiltonian2}, we can utilize the full existing machinery for the electronic case. In particular, this means that we have now polaritonic orbitals $\phi_i'(\bz \sigma)$ as fundamental entities that have as coordinate $\bz \equiv \br \boldsymbol{q}$, where $\boldsymbol{q}$ is an $M$-dimensional (number of photon modes) vector. Additionally, the one- and two-body terms are replaced by their polaritonic counterparts, i.e.\ $(t,v,w)\rightarrow (t',v',w')$ as given in Eqs.~\eqref{eq:dressedkinetic}, \eqref{eq:dressedpotential} and \eqref{eq:dressedinteraction}. We can then transform straightforwardly the energy expression of a given electronic-structure theory into a polariton energy expression $E_m^{t,v,w}[\Psi]\rightarrow E_m^{t',v',w'}[\Psi']$, because the connection between $E_m$ and $\hat{H}_m$ is defined by the one- and two-body terms and the particle number alone. Also the constraints directly transfer to the polariton system, leading to the Lagrangian term $\mathcal{C}[\Psi,\{\lambda_k\}] \to \mathcal{C}[\Psi',\{\lambda_k\}]$. Lastly, since polaritons are particles with a more complicated hybrid statistics than electrons (see Sec.~\ref{sec:dressed_orbitals}), we need to add to the Lagrangian a further constraint term $\mathcal{G}[\gamma_e]$\footnote{The precise form of this term depends on the method that is used. In the last part of this section, we give an explicit example in Eq.~\eqref{eq:constraints_inequality_augmented_lagrangian}.} to enforce the constraints
\begin{align}
\label{eq:ConstraintsInequality}
g_i[\gamma_e]= 1 - n_i^e \geq 0 \quad\forall\,i = 1,\dotsc,N.
\end{align}
With this definition, the energy expression $E_m^{t',v',w'}[\Psi']$ and the constraints, we are now able to generalize the minimization problem of Eq.~\eqref{eq:MinimizationProblemEST} to the \emph{generic polaritonic minimization problem}
\begin{alignat}{2}
\label{eq:MinimizationProblem}
&\text{minimize }\, &&E_m^{t',v',w'}[\Psi'] \nonumber\\
&\text{subject to }\quad &&c_k[\Psi'] = 0 \\
& &&g_i\bigl[\gamma_e[\Psi']\bigr] \geq 0.		\nonumber
\end{alignat}
Since there are many possible strategies to solve the minimization problem \eqref{eq:MinimizationProblem} and a good choice might depend on the specific electronic-structure theory that is considered, we finish the general discussion here. Instead, we conclude the section with a concrete example.
We will apply the above rules to HF theory, which leads to polaritonic HF theory. This means that we approximate the density matrix of the exact dressed wave function of Eq.~\eqref{eq:DressedManyBodyWF} by the density matrix of a single Slater determinant with 
orbitals ${\phi}'_1,\dotsc, {\phi}'_N$, i.e. 
\begin{align*}
\Phi'(\bz_1 \sigma_1,\dotsc,\bz_N \sigma_N)=
\frac{1}{\sqrt{N!}}\sum_{\pi_j\in P_N} (-1)^{j}{\phi}'_{\pi_j(1)}(\bz_1 \sigma_1)\dotsb{\phi}'_{\pi_j(N)}(\bz_N \sigma_N),
\end{align*}
where $P_N$ denotes the permutation group on $N$ elements and the index $j$ is chosen such that it is even (odd) for an even (odd) permutation $\pi_i \in S_N$. Further, we consider a spin-restricted formalism, i.e.\ we assume that the number of electrons $N$ is even and define ${\phi}'_{2k-1}(\bz \sigma)=\phi'_k(\bz)\alpha(\sigma), {\phi}'_{2k}(\bz \sigma)=\phi'_k(\bz)\beta(\sigma)$ for $k=1,\dotsc,N/2$, where $\alpha,\beta$ are the usual spin-orbitals. We again note that we do not necessarily enforce with our constraints that the auxiliary Slater determinant has the right symmetry but rather its 1RDM. In this regard polaritonic HF becomes actually a 1RDM functional theory for polaritonic problems rather than a wave-function based method~\cite{Buchholz2019}. With this ansatz, we calculate the energy expectation value for the Hamiltonian of Eq.~\eqref{AuxiliaryHamiltonian2}, which reads 
\begin{align}
\label{eq:PdHF_Energy}
E_{HF}'=2 \sum_i \braket{\phi_i' | (\hat{T}[t'] + \hat{V}[v']) \phi_i'} + \sum_{i,k} \left[2 \braket{\phi_k' | \hat{J}'_i[w']\phi_k'} - \braket{\phi_k' | \hat{K}'_i[w']\phi_k'}\right],
\end{align}
where we introduced the ``dressed'' Coulomb-operator $\hat{J}_i'$ which acts as
\begin{subequations}
\begin{align}
\hat{J}_i'\phi_k'(\br \bq) &= \int\td\bz' \phi_i'{}^*(\bz')w'(\bz;\bz') \phi_i'(\bz') \phi'_k(\bz) \\
\intertext{and the ``dressed'' exchange-operator $\hat{K}_i'$ that acts as\cite{Szabo2012}}
\hat{K}_i'\phi_k'(\bz) &= \int\td\bz' \phi_i'(\bz)w'(\bz;\bz')\phi_i'{}^*(\bz') \phi_k'(\bz').
\end{align}
\end{subequations} 
The polaritonic one- and two-body terms are given by~\eqref{eq:dressedkinetic}, \eqref{eq:dressedpotential} and~\eqref{eq:dressedinteraction}, respectively. With this we find that $E_{HF}'=E_{HF}(t',v',w',\{\phi_k'\})$. Consequently, we also find structurally the same derivative (that for HF theory is called the Fock-matrix), which reads
\begin{align}
\label{eq:PdHF_Fockmatrix}
\nabla_{\phi_k^*} E_{HF}'= \hat{H}^1{}' \phi_k' = 2 (\hat{T}[t']+\hat{V}[v']) \phi_k' + 2 \sum_{i} \left[2  \hat{J}_i'[w']\phi_k' - \hat{K}_i'[w']\phi_k'\right].
\end{align}
Since we consider only one Slater determinant, the orbitals $\phi_i'$ are also the eigenfunctions of the system's dressed 1RDM $\gamma$. Because of the spin-restriction, it suffices also to consider the spin-summed version $\gamma(\bz,\bz')=2 \sum_{i=1}^{N/2} \phi_i'{}^*(\bz') \phi_i'(\bz)$, which we denote with the same symbol.
We see that $\gamma[\Phi']$ has occupations (eigenvalues) of 2 instead of 1 because of the spin-summation. This transfers to the natural occupation numbers of the electronic 1RDM 
\begin{align*}
\gamma_e(\br,\br')= \int\td\bq \gamma(\br \bq,\br' \bq).
\end{align*}
Now, we have defined all the terms that enter the minimization problem \eqref{eq:MinimizationProblem} and we can proceed with demonstrating how to solve it. Algorithmic-wise, we are confronted with enforcing the additional inequality constraints \eqref{eq:ConstraintsInequality} in extension to the original (HF) minimization problem in~\eqref{eq:MinimizationProblemEST}.
We start by noting that the constraint functions $g_i$ depend on $\gamma\{\phi_k'\}$, which can be directly calculated from the polariton orbitals,
via the eigendecomposition of $\gamma_e$.
Since the diagonalization of $\gamma_e$ is a non-trivial step for large systems (or in real-space) and thus can be a bottleneck of the minimization, it is helpful to consider natural and dressed orbitals as \emph{independent} variables of the minimization and enforce their connection as an additional constraint.\footnote{This is similar to considering $\phi$ and $\phi^*$ as independent.} We thus define $g_i=g_i[\gamma_e\{\phi_k', \psi^{e}_i\}]$ and include the necessary orthonormality of the $\psi^{e}_i$ by a third set of conditions
\begin{align}
\label{eq:ConstraintsNO}
f_{ij}=\braket{\psi^{e}_i|\psi^{e}_j} -\delta_{ij}=0,
\end{align}
that we include in the minimization by a third Lagrange-multiplier term $-\sum_{ij}\bar{\theta}_{ij} f_{ij}$. Note that this construction automatically linearizes the constraints \eqref{eq:ConstraintsInequality} during one minimization step, where the $\psi^e_i$ are fixed.

To enforce now these inequality constraints, we use an \emph{augmented Lagrangian} algorithm, following the book of \citet{Nocedal2006}, Ch.~17.3.\footnote{Note that the algorithm is presented there for equality constraints. However, in the beginning of Ch.~17, the authors remark on how to generalize the method for inequality constraints.} 
We chose this algorithm, since it simply extends a given Lagrangian with penalty terms. Hence, we can make use of any existing implementation that solves the minimization problem of Eq.~\eqref{eq:MinimizationProblemEST} and just add the extra terms with corresponding extra iteration loops. To test this, we conducted all the numerical examples of Sec.~\ref{sec:study_case} with a standard electronic-structure algorithm~\cite{Payne1992}, which we extended by the augmented-Lagrangian method for the inequality constraints. This extension involves two extra terms. A linear (so-called augmented) term, $-\sum_i \nu_i g_i$ with Lagrange-multipliers $\nu_i$ that are initialized to zero and updated to values $\nu_i>0$ only if the minimization reaches the corresponding boundary of the feasible region where $g_i=0$. And a second non-linear term, that adds a penalty function $P=\mu / 2 \sum_i ([g_i]^-)^2$, where $[y]^-$ denotes max$(-y, 0)$, which penalizes violations of condition \eqref{eq:ConstraintsInequality} quadratically, but has no effect in the so-called \emph{feasible} region of configuration space, where the conditions \eqref{eq:ConstraintsInequality} are satisfied.

Specifically for our example, the extra Lagrangian term of the translation rules depicted in Fig.~\ref{fig:polariton_construction} is given by
\begin{align}
\label{eq:constraints_inequality_augmented_lagrangian}
\mathcal{G}[\gamma_e\{\phi_k',\psi^{e}_i\}]= - \sum_i \lambda_i g_i[\gamma_e\{\phi_k',\psi^{e}_i\}] + \mu \sum_i ([g_i]^-[\gamma_e\{\phi_k',\psi^{e}_i\}])^2 - \sum_{ij} \bar{\theta}_{ij} f_{ij} [\gamma_e\{\psi^{e}_i\}].
\end{align}
The full Lagrangian for the polaritonic HF minimization problem reads then
\begin{align}
\label{eq:Lagrangian}
L_{HF}'[\gamma\{\phi_k',\psi^{e}_i\}]=&	E_{HF}' -\sum_{ij} \bar{\epsilon}_{ij} h_{ij}[\gamma\{\phi_k'\}] +\mathcal{G}[\gamma_e\{\phi_k',\psi^{e}_i\}]
\end{align}
and the corresponding first order conditions for a minimum (stationary point) of $L_{HF}'$ are
\begin{subequations}
	\begin{align}
	\label{eq:GradientPhi}
	0=&\nabla_{\phi_k'{}^*} L_{HF}'= \hat{H}^1\phi_k' - \sum_j\bar{\epsilon_{kj}} \phi_j' + \sum \left[ \lambda_i -\mu [g_i]^-\right]  \hat{G}_i\phi_k'\\
	\label{eq:GradientPsiGamma}
	0=&\nabla_{\psi^{e}_i{}^*}L_{HF}' = (\mu [g_i]^- -\lambda_i )\int\td\br' \gamma_e(\br',\br)\psi^{e}_i(\br') - \sum_j \bar{\theta}_{ij} \psi^{e}_j,
	\end{align}
\end{subequations}
where we considered $\phi_k'$ and $\phi_k'{}^*$ as independent and defined $\hat{G}_i\phi_k'(\br \bq)=n_k^e\int\td\br' \psi^{e}_i{}^*(\br' )\phi_k'(\br' \bq)$ $ \psi^{e}_i(\br \sigma)$.
Additionally, we can diagonalize the Lagrange-multiplier matrices $\bar{\epsilon}_{ij}=\delta_{ij}\epsilon_j$ and $\bar{\theta}_{ij}=\delta_{ij}\theta_j$, since the orbital-dependent Hamiltonian $\hat{H}^1$ and the electronic 1RDM $\gamma_e$ are hermitian. We also want to remark on the second gradient equation, c.f.\ Eq.~\eqref{eq:GradientPsiGamma}, which is much simpler than it looks like on a first glance. In fact, solving Eq.~\eqref{eq:GradientPsiGamma} is equivalent to solving first the eigenvalue equation for $\gamma_e$ (see the paragraph above Eq.~\eqref{eq:Nrepresentability}) and then replacing $\theta_i=n_i^e(\mu [g_i]^- -\lambda_i)$. With these definitions, we are able to perform polaritonic HF calculations by numerically solving the Eqs.~\eqref{eq:GradientPhi} and \eqref{eq:GradientPsiGamma} with the expressions \eqref{eq:PdHF_Energy} and \eqref{eq:PdHF_Fockmatrix}. We did this for a model Hamiltonian and will present the results in the next section.

\vspace{1cm}
Before we come to the results, we want to conclude this section with a brief remark on how to generalize the just presented example of polaritonic HF theory. Since we consider in HF theory only one Slater determinant, the equations become especially simple. However, including (polaritonic) correlation beyond this approximation is easily possible by, for instance, polaritonic QEDFT. To do so, we can leave the structure of the implementation exactly as it is, but just exchange the operators $\hat{K}_i'$ in \eqref{eq:PdHF_Energy} and \eqref{eq:PdHF_Fockmatrix} by the exchange-correlation functional of our choice. Another simple extension is polaritonic 1RDM functional theory\cite{Buchholz2019}. Additionally, we want to mention that the conditions \eqref{eq:ConstraintsInequality} can be expressed more generally as $1-\hat{\gamma}_e'\succcurlyeq 0$, where the $\succcurlyeq$ symbol denotes \emph{positive-semidefiniteness}, i.e. for any single-particle wave function $\psi$ it holds that $1 - \braket{\psi |\hat{\gamma}_e \psi}\geq 0$. This general form is exploited by \emph{semi-definite programming} methods, which might be very efficient for the current problem.
\section{Exemplification: polaritonic Hartree--Fock theory on a lattice}
\label{sec:study_case}
In this section, we demonstrate numerically how to employ the dressed approach with the correct hybrid statistics for the example of (spin-restricted) HF theory. We want to stress here that from a computational perspective, enforcing the hybrid statistics \emph{does not increase} the scaling of the method. Still, the extra inequality conditions, c.f. \eqref{eq:ConstraintsInequality}, introduce additional iteration loops and thus more steps are required until convergence. But each of these steps is less expensive than the calculation of the gradient (which is the Fock-matrix in HF). A typical HF solver for a matter system, e.g. that diagonalizes the Fock-matrix self-consistently, scales as $B_m^3$ where $B_m$ is the size of the basis that is used. Polaritonic HF has exactly the same scaling with the basis size, but of course we have to consider a larger basis. If the photon mode(s) are satisfactorily described by a basis of size $B_{ph}$ and the matter-part of the system again with a basis of size $B_m$ than polaritonic HF scales as $(B_m*B_{ph})^3$. Thus, we get a factor of $B_{ph}^3$ in addition to the matter description. Note that this holds for 1D, 2D and 3D systems equivalently.\footnote{A similar consideration holds also for a real-space description, where the Fock-matrix cannot be explicitly constructed, but is diagonalized iteratively by, e.g., a conjugate-gradient algorithm like the one of our implementation\cite{Payne1992}. The scaling of such a method in the electronic case is of the order $O(B_m \ln B_m N^2)$ and can even be reduced with state-of-the-art algorithms to $O(B_m \ln B_m N)$\cite{Lin2016}, where N is the number of electrons. Such methods thus scale better than a direct diagonalization of the Fock-matrix, but at the same time the underlying basis size $B_m$ describes the number of grid points and thus is typically much larger than $B_m$ in orbital-based codes. However, for very large systems, both descriptions can have comparable $B_m$. A real-space like description of the displacement coordinates of the photon modes would in a similar manner be quite inefficient for a few photons, but might become even advantageous for large photon numbers.}

However, for the exemplification, we consider a one-dimensional lattice system  that couples to one photon mode in dipole approximation with frequency $\omega$ (a sketch of the setup has been depicted in Fig. \ref{fig:cavity_sketch}). We want to remark that including more modes is in principle possible (see Sec. \ref{sec:est}), but the one-mode case is the standard. The Hamiltonian is of the form of Eq.~\eqref{eq:Hamiltonian} and reads 
\begin{align}
    \hat{H}=\overbrace{-t\sum_{i}^{B_m} \sum_{\sigma=\uparrow, \downarrow} (\hat{c}^{\dagger}_{i,\sigma}\hat{c}_{i+1,\sigma} + \hat{c}^{\dagger}_{i+1,\sigma}\hat{c}_{i,\sigma}) +\sum_{i=1}^{B_m}v_i \hat{n}_i}^{\hat{H}_m} + \overbrace{ \frac{\lambda^2}{2} \sum_{i,j=1}^{B_m} \hat{n}_i \hat{n}_j x_i x_j }^{\hat{H}_d}\nonumber\\
    \underbrace{- \sqrt{\frac{\omega}{2}}(\hat{a}^{\dagger}+\hat{a})\lambda \sum_{i=1}^{B_m} x_i\hat{n}_i }_{\hat{H}_I}
   +\underbrace{\omega (\hat{a}^+\hat{a} +1/2)}_{\hat{H}_{ph}},
   \label{eq:Hamilt_example}
\end{align}
with hopping $t=\frac{1}{2 \Delta x^2}$ corresponding to a second-order finite difference approximation for a grid with spacing $\Delta x$, where we choose $t=0.5$ for all calculations, which corresponds to a spacing between neighbouring sites of $\Delta x=1$ bohr and a local scalar potential with value $v_i$ on site $i$. We have set the Coulomb repulsion to zero in this example to highlight the influence of the matter-photon coupling and how well the polaritonic HF approach can capture it. Nevertheless, due to the dipole self-energy term $\hat{H}_d$ we have a mode-induced dipole-dipole interaction among the electrons. This type of interaction is important in many fundamental quantum-optical questions, such as the quest for a super-radiant phase in the strong-coupling case\cite{Dzsotjan2011,Griesser2016,Plankensteiner2017}.
Further, the electron basis $B_m$ is determined by the number of sites, $x_i=i-x_0$ is the position with respect to the middle of our lattice $x_0$, $\hat{c}^{(\dagger)}_{i,\sigma}$ are the fermionic creation (annihilation) operators that satisfy the anticommutation relation $[\hat{c}_{i,\sigma},\hat{c}^{\dagger}_{j,\sigma'}]_{+}=\delta_{ij}\delta_{\sigma \sigma'}$, and $\hat{n}_i=\hat{c}^{\dagger}_{i,\uparrow} \hat{c}_{i,\uparrow}+\hat{c}^{\dagger}_{i,\downarrow} \hat{c}_{i,\downarrow}$ is the density operator. We have chosen a simple lattice model, since this allows us to have still exact numerical reference data to compare to. In contrast to standard electronic-structure problems, there are are no (numerically exact) references solutions currently available for real three-dimensional systems in the continuum. To the best of our knowledge there are only QEDFT simulations (at several levels of approximations) available for realistic three-dimensional systems~\cite{Flick2018a, Flick2019}.

For the implementation and to go to the polariton picture, we express the matter $\hat{H}_m$ plus dipole part $\hat{H}_{d}$ of our Hamiltonian in matrix form by using the basis states $\ket{\tilde{\psi}_{i,\sigma}}=\hat{c}^{\dagger}_{i,\sigma}\ket{0}$. As a basis for the photon subsystem, we utilize the eigenstates $\chi_i$ of the photon energy operator, i.e.\ $\hat{H}_{ph}\chi_{\alpha}=(\alpha + 1/2) \chi_{\alpha}$, which are photon number states. To calculate the coupling term of the energy expression $H_I$, c.f. Eq.~\eqref{eq:PdHF_Energy}, we express the displacement operator $p_{\alpha}=\frac{1}{\sqrt{2\omega_{\alpha}}}(\hat{a}^{\dagger}_{\alpha}+\hat{a}_{\alpha})$ in this basis as well. 
To then construct the auxiliary Hamiltonian $\hat{H}'$ to Eq.~\eqref{eq:Hamilt_example} according to the rules from Sec.~\ref{sec:est}, we would need to define  the auxiliary terms $t',v',w'$, c.f. Eqs. \eqref{eq:dressedkinetic}-\eqref{eq:dressedinteraction}. Since in this section we employ a second-quantized picture, it is less convenient to define these kernel-like quantities, but directly the many-body operators $\hat{T}[t'], \hat{V}[v'], \hat{W}[w']$. For the one-body terms $T[t']+V[v']$, this is straightforward and the expression reads 
\begin{align}
\label{eq:TB_Onebody}
T[t']+V[v'] =& \hat{H}_m + \hat{H}_{ph} + \frac{\lambda^2}{2} \sum_{i=1}^{B_m} \hat{n}_i \hat{n}_i x_i^2- \sqrt{\frac{\omega}{N}}(\hat{a}^{\dagger}+\hat{a}) \lambda \sum_{i=1}^{B_m} x_i\hat{n}_i.
\end{align}	
However, the interpretation of the operators is different from before, because we have to apply them to polaritonic basis states. Since we consider the spin-restricted formalism introduced in the end of Sec.~\ref{sec:est}, we neglect the spin-dependency of the electronic part of the basis $\tilde{\psi}_{i,\sigma}\rightarrow\psi_{i}$ and define $\ket{\phi'_{i\alpha}}=\ket{\psi_{i}\chi_{\alpha}}$. We can then derive the kernel expression $(t+v)(x,q) \rightarrow(t'+v')_{i\alpha}^{j\beta}=\braket{\psi_{i}\chi_{\alpha}|(\hat{T}' + \hat{V}')\psi_{j}\chi_{\beta}}$ as matrix elements
\begin{align}
	(t'+v')_{i\alpha}^{j\beta}=& -t(\delta_{i,j+1}+\delta_{i+1,j}) + v_i \delta_{ij} + \omega (\delta_{\alpha,\beta}+\tfrac{1}{2}) + \tfrac{\lambda^2}{2}(x_i-x_0)^2 \delta_{ij} \delta_{\alpha,\beta} \nonumber\\
	&- \lambda\sqrt{\tfrac{\omega}{N}} (x_i-x_0) \delta_{ij} (\sqrt{\beta+1}\delta_{\alpha,\beta+1}+\sqrt{\beta}\delta_{\alpha,\beta-1}).
\end{align}
For the two-body term, a definition analogously to \eqref{eq:TB_Onebody} is more difficult, since we have to differentiate the two polaritonic coordinates. For the sake of the analogy, we formally write
\begin{align}
	\hat{W}'=&\lambda^2 \sum_{i^1\neq j^2=1}^{B_m} \hat{n}_{i^1} \hat{n}_{j^2} x_{i^1} x_{j^2}- 
	\sqrt{\frac{\omega}{N}}(\hat{a}^2{}^{\dagger}+\hat{a}^2)\lambda \sum_{i^1=1}^{B_m} x_{i^1}\hat{n}_{i^1}- 
	\sqrt{\frac{\omega}{2}}(\hat{a}^1{}^{\dagger}+\hat{a}^1)\lambda \sum_{^2i=1}^{B_m} x_{i^2}\hat{n}_{i^2},
\end{align}
where the upper indices differentiate the two polaritonic orbitals that both have an electronic and photonic part. This is to be understood in the following sense: To define the corresponding kernel $w(x^1,q^1,x^2,q^2)\rightarrow w^{i^1\alpha^1i^2\alpha^2}_{j^1\beta^1j^2\beta^2}= \braket{\psi_{i^1}\chi_{\alpha^1}\psi_{i^2} \chi_{\alpha^2}|\hat{W}\psi_{j^1}\chi_{\beta^1}\psi_{j^2}\chi_{\beta^2}}$, the operators only act on the basis elements with the same indices. The kernel $(w_{\text{self}})^{i^1\alpha^1i^2\alpha^2}_{j^1\beta^1j^2\beta^2}$ for the self-interaction part $\hat{W}_{\text{self}}=\lambda^2 \sum_{i^1\neq j^2=1}^{B_m} \hat{n}_{i^1} \hat{n}_{j^2} x_{i^1} x_{j^2}$ reads for example
\begin{align}
	(w_{\text{self}})^{i^1\alpha^1i^2\alpha^2}_{j^1\beta^1j^2\beta^2}
	=&\lambda^2 (x_{i^1}-x_0) (x_{i^2}-x_0) \delta_{i^1j^1} \delta_{i^2j^2}  \delta_{\alpha^1\beta^1} \delta_{\alpha^2\beta^2}.
\end{align}
With these definitions, we can calculate the polaritonic HF energy expression, c.f.~\eqref{eq:PdHF_Energy} and the polaritonic HF Fock-matrix, c.f.~\eqref{eq:PdHF_Fockmatrix}. Then, we employ the augmented Lagrangian algorithm as discussed in Sec. \ref{sec:est} to find the polaritonic HF ground state of the model system.

As a first example, we illustrate the violation of the Pauli principle if we do not enforce the right symmetries, as also discussed in Sec.~\ref{sec:dressed_orbitals} for a simple uncoupled problem. 
To this end, we compare ground-state energies, electronic 1RDMs and the photon number of a small 4-electron system obtained with the two different HF ground states, i.e.\ polaritonic HF using density matrices with the exact symmetry, c.f.~\eqref{eq:SymmetryPauliDressedFermion}, and polaritonic HF with only fermionic symmetry (which we call in this section fermionic HF). We can expect deviations between both polaritonic-HF theory levels for systems that contain more than one orbital. In our spin-restricted case this corresponds to more than two electrons and that is why we chose here $N=4$.
Further we set the external potential to zero, i.e.\ $v_i=0 \,\forall i$. Since we need to calculate the exact coupled electron-photon many-body ground state from a configuration space that grows exponentially fast with the size of the basis sets and the electron number, we choose a small box of length $L=5$ bohr. This corresponds to a matter basis of $B_{m}=6$ spatial sites times two spin states for each electron. For the photon-subsystem, we consider $B_{ph}=5$ photon number states for which all relevant quantities are sufficiently converged.\footnote{For example, deviations in the energy or photon number between $B_{ph}=5$ and $B_{ph}=6$ are maximally of the order of $10^{-4}$.} Despite the small basis sets and electron number employed, the many-body configuration space has the considerable size of $(2B_{m})^{N}*B_{ph}\approx 10^5$, which is already at the edge of standard exact diagonalization solvers: matrices of this size can still be diagonalized without special efforts like parallelization. Since we only aim for a benchmark study here, this limitation is not problematic, but it shows how expensive the exact solutions of coupled electron-photon systems computationally are. The need for numerically manageable approximations is evident here.

\begin{figure}[H]
	\centering
	\begin{overpic}[width=0.49\columnwidth]{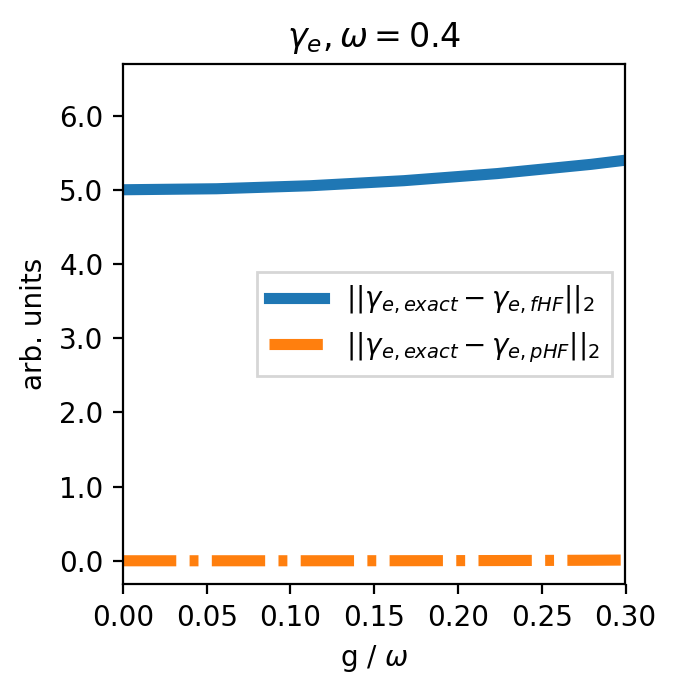}
		\put (22,83) {\textcolor{black}{(a)}}\hfill
	\end{overpic}
	\hfill
	\begin{overpic}[width=0.49\columnwidth]{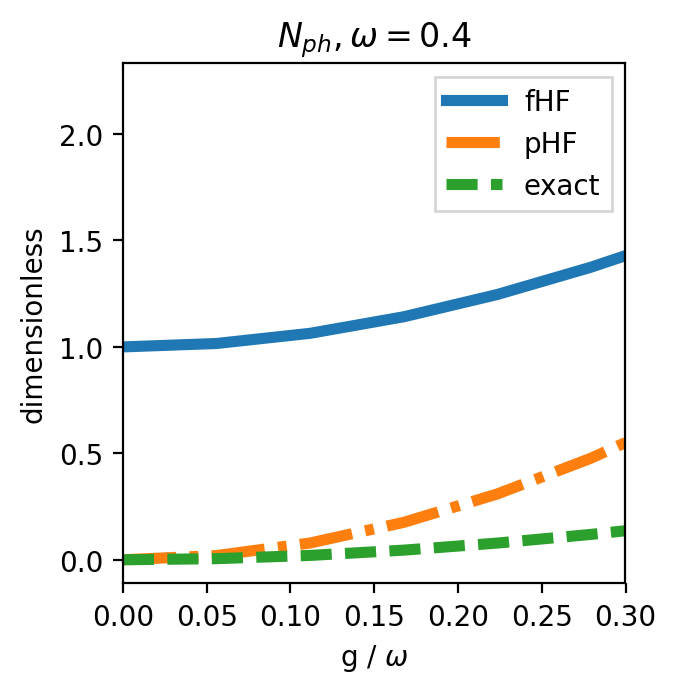}
		\put (22,83) {\textcolor{black}{(b)}}\hfill
	\end{overpic}
	\caption{Comparison of the electronic 1RDM $\gamma_e$ (a) and the photon number $N_{ph}$ (b) for the 4-electron system with $\omega = 0.4 $ hartree for varying coupling strength $g/\omega$. In (a) the norm difference between the exact 1RDM and the polaritonic HF (pHF) 1RDM (dashed-dotted orange line) and between the exact 1RDM and fermionic HF 1RDM (fHF) (solid blue line) are displayed. In (b) the exact photon number (dashed green line) and the polaritonic HF (dashed-dotted orange line) and fermionic HF (solid blue line) photon numbers are shown. In both cases, fermionic HF deviates much stronger from the exact reference than polaritonic HF due to the wrong symmetry.}
	\label{fig:tb_gamma_nphot}
\end{figure}

We first compare the electronic 1RDMs $\gamma_e$, c.f. Eq.~\eqref{eq:GammaElectronic}, and the photon numbers $N_{ph}=\braket{\hat{N}_{ph}}$, c.f. Eq.~\eqref{eq:PhotonNumber} using the connection formula of Eq.~\eqref{eq:PhotonEnergyConnection}, for varying coupling strengths $g/\omega =\lambda/\sqrt{2\omega}$ and $\omega=0.4$. We choose these quantities because they provide us with a consistent measure of how well the electronic part and the photonic part of the system are approximated. The electronic 1RDM determines all electronic one-body quantities and is therefore a very precise measure of the quality of the approximation in the electronic sector. Indeed, the photonic 1RDM in the single mode case corresponds to the photon number $N_{ph}$, which is an equivalently good measure for the quality in the photonic sector. In Fig.~\ref{fig:tb_gamma_nphot} (a), we display the difference of the exact electronic 1RDM from the one of the polaritonic HF and fermionic HF approximations, measured by the Frobenius norm $\lVert A \rVert_2=\sqrt{\sum_{ij} A_{ij}^2}$ for a matrix $A_{ij}$. We see that for all coupling strengths the polaritonic HF 1RDM (dashed-dotted orange line), which enforces the right hybrid statistics, remains very close to the exact solution indicating that the electronic subsystem is captured very well within this approximation. The fermionic HF (solid blue line) approximation, however, deviates strongly due to its wrong purely fermionic character. 

The same behavior is also encountered in the photonic subsector, where in Fig.~\ref{fig:tb_gamma_nphot} (b) the photon number of the exact calculation (dashed green line) is compared to the polaritonic HF (dashed-dotted orange line) and to the fermionic HF photon number (solid blue line). We therefore find, similarly to the simple uncoupled problem in Sec.~\ref{sec:dressed_orbitals} (for $g/\omega=0$ we recover this case exactly), that for the same energy expression using the wrong statistics leads to sizeable errors.

\begin{figure}[H]
	\centering
		\begin{overpic}[width=0.49\columnwidth]{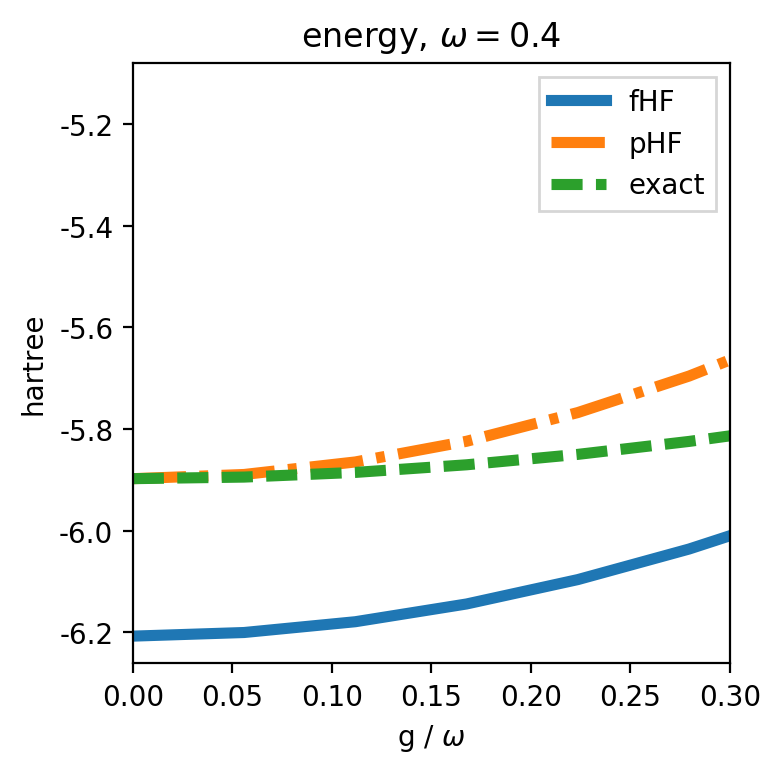}
		\put (22,83) {\textcolor{black}{(a)}}\hfill
	\end{overpic}
	\hfill
	\begin{overpic}[width=0.49\columnwidth]{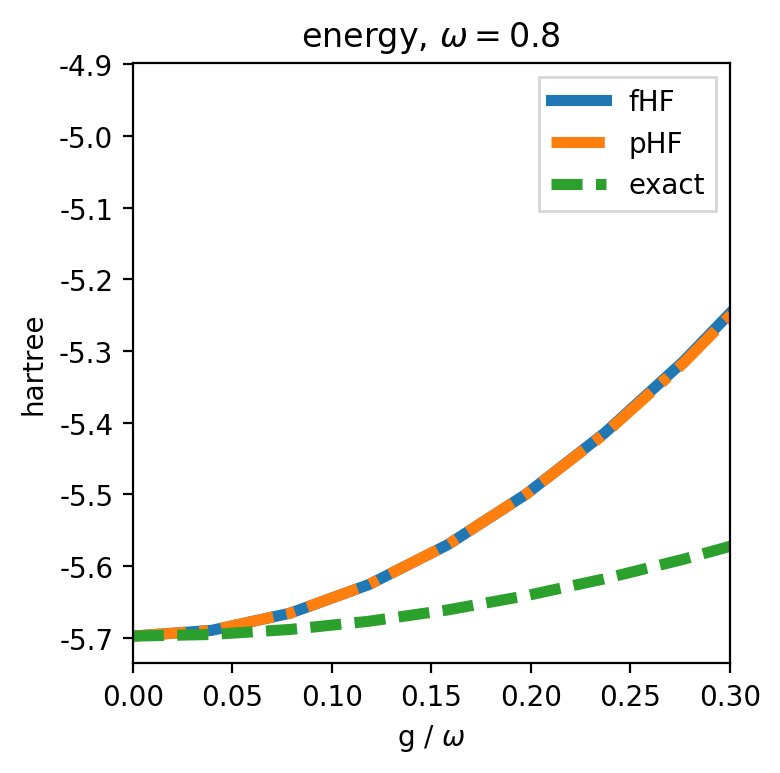}
		\put (22,83) {\textcolor{black}{(b)}}\hfill
	\end{overpic}
	\caption{Total energy for the 4-electron system with $\omega = 0.4 $ hartree (a) and $\omega = 0.8 $ hartree (b) for varying coupling strength $g/\omega$. While in (a) the fermionic HF approximation (solid blue line) can achieve unphysically low energies when compared to the exact solution (dashed green line) due to the wrong statistics, in (b) the minimal-energy condition singles out the right statistics without further constraints. The polaritonic HF (dashed-dotted orange line) by construction always has the right hybrid statistics and thus is variational, i.e.\ the energy is always above the exact energy.}
	\label{fig:tb_energy}
\end{figure}

The same problem is encountered also in Fig.~\ref{fig:tb_energy} (a), where we display the total energy $E=\braket{\hat{H}}$ of the coupled system as a function of the coupling strength $g/\omega$. While the polaritonic HF (dashed-dotted orange line) is variational, i.e.\ due to the right statistics we are always equal or above the exact energy (dashed green line), the fermionic HF (blue solid line) breaks the proper symmetry and thus can reach energies below the physically accessible ones. However, again in close analogy to the uncoupled example in Sec.~\ref{sec:dressed_orbitals}, if we increase the frequency of the photon field such that it is much more costly to excite photons than electrons, the minimal-energy conditions can single out the correct statistics, as displayed in Fig.~\ref{fig:tb_energy}(b). That is, for $\omega$ large enough the constraints $g_i[\gamma\{\phi_k', \psi_i^e \}] \geq 0$ of Eq.~\eqref{eq:MinimizationProblem} are trivially fulfilled.  

\begin{figure}[H]
	\centering
	\begin{overpic}[width=0.49\columnwidth]{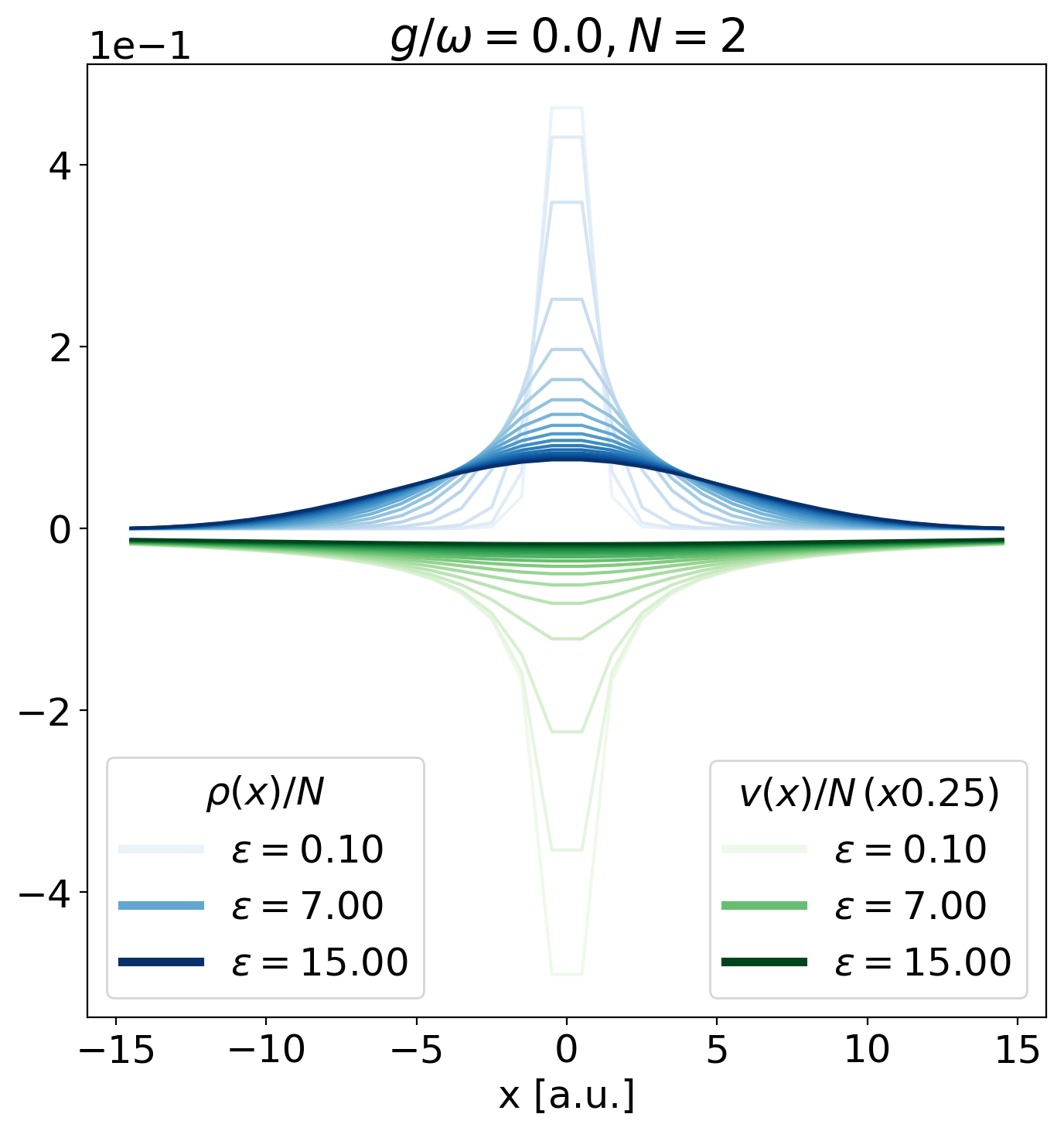}
		\put (13,85) {\textcolor{black}{(a)}}
	\end{overpic}
	\hfill
	\begin{overpic}[width=0.49\columnwidth]{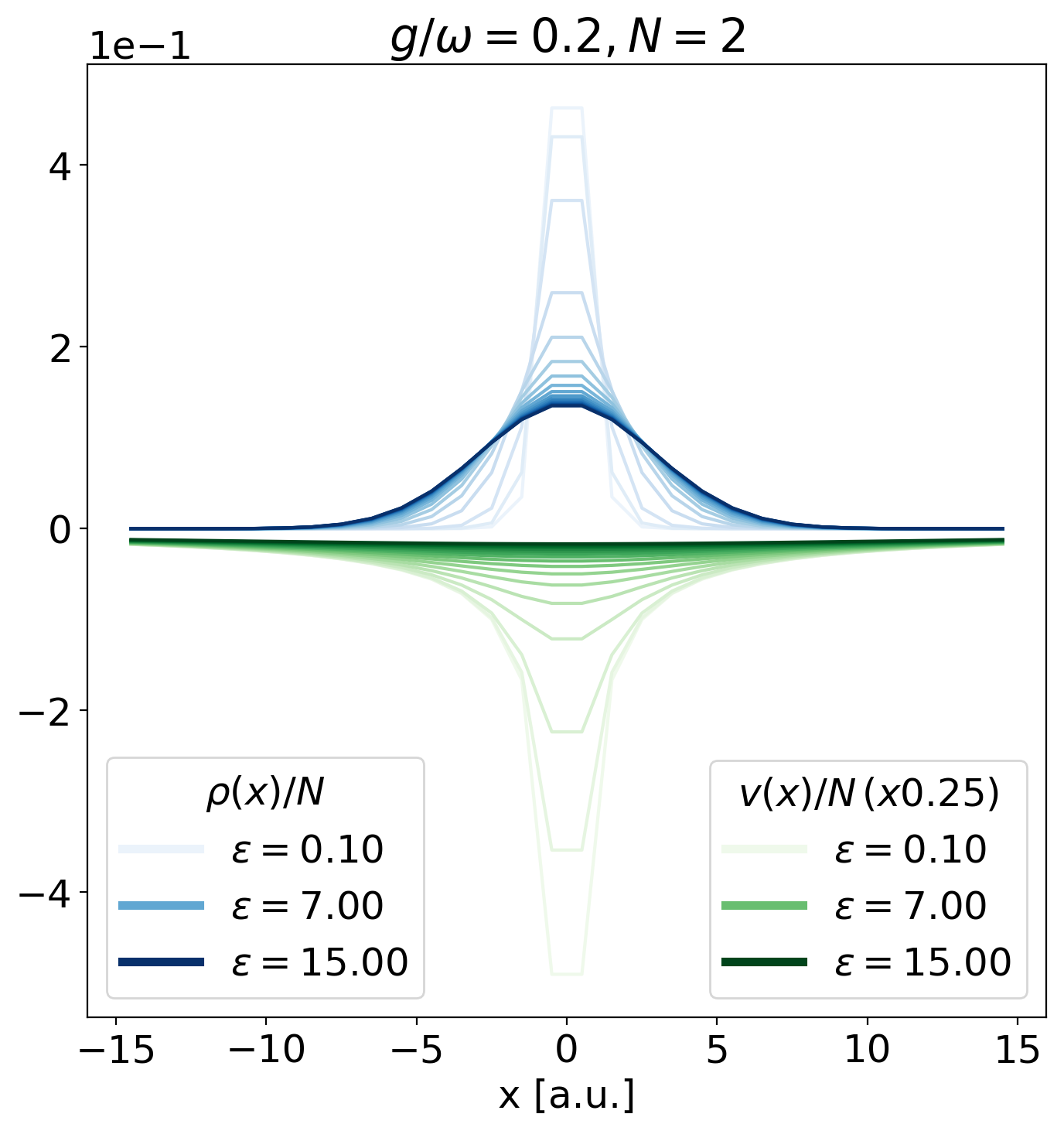}
		\put (13,85) {\textcolor{black}{(b)}}
	\end{overpic}
	\hfill
	\begin{overpic}[width=0.49\columnwidth]{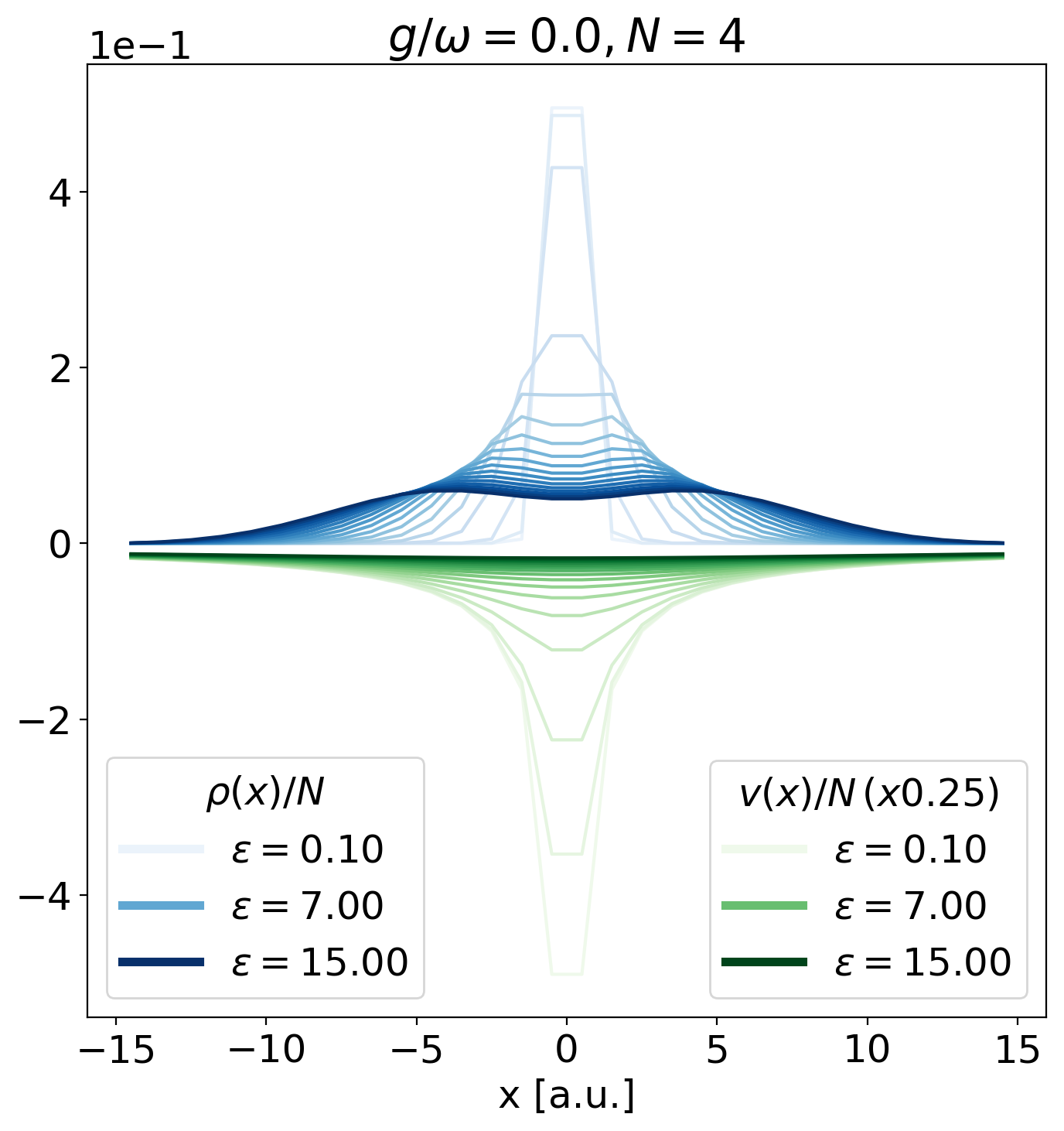}
		\put (13,85) {\textcolor{black}{(c)}}
	\end{overpic}
	\hfill
	\begin{overpic}[width=0.49\columnwidth]{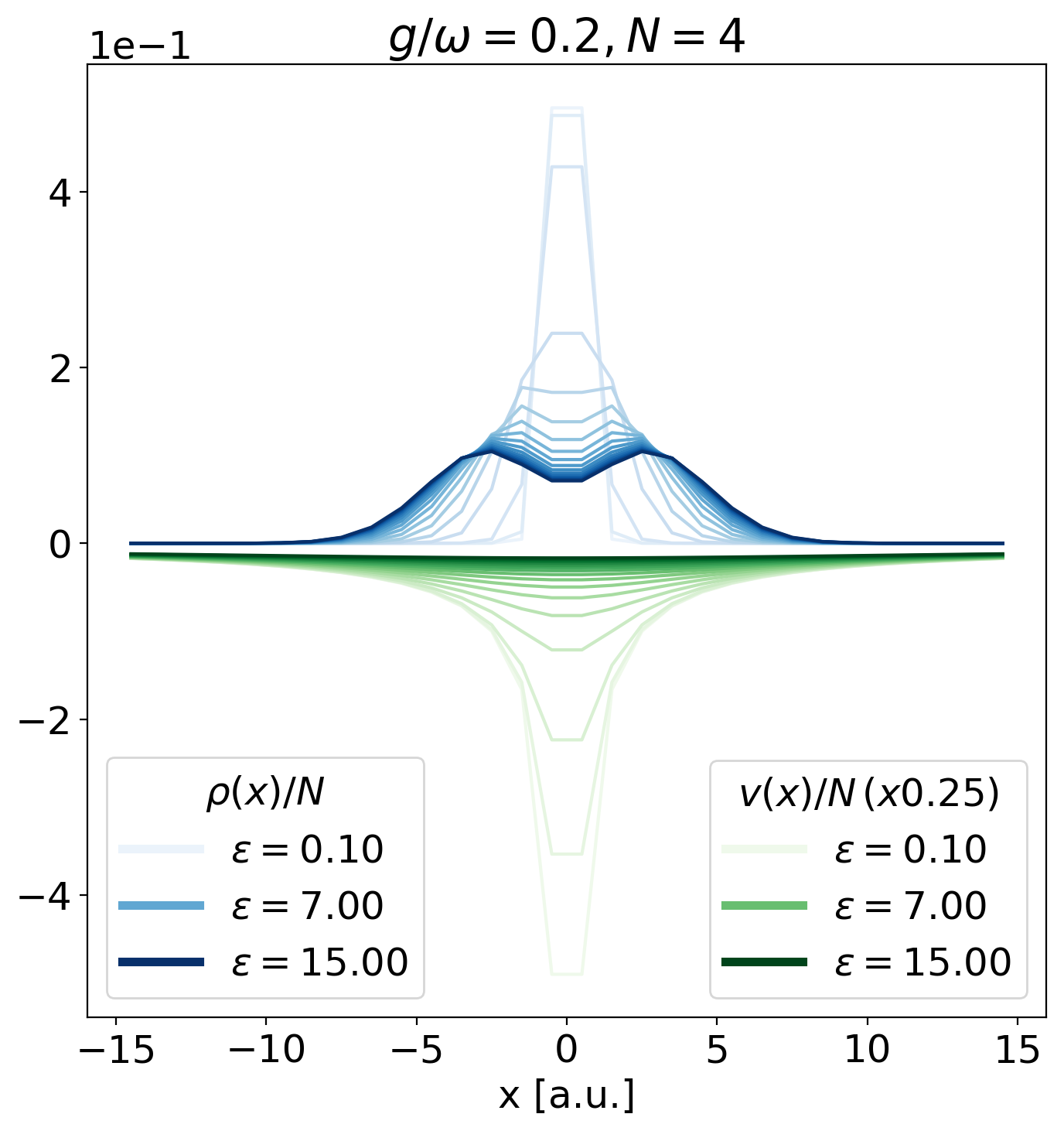}
		\put (13,85) {\textcolor{black}{(d)}}
	\end{overpic}
	\caption{Every plot depicts the normalized electron density $\rho(x)/N$(blue) with corresponding local potentials $v(x)/N=1/\sqrt{x^2 - \epsilon^2}$ (green, rescaled by a factor of $0.25$ for better visibility) for a series of softening parameters $\epsilon$ of the 2- (upper row) and 4-electron (lower row) system and for coupling strength $g/\omega=0$ (left column) and $g/\omega=0.2$ (right column). We see how different both systems respond to the coupling depending on the degree of confinement, measured by $\epsilon$. Note that the legends only depict three example lines (the smallest, largest and middle values of $\epsilon$), respectively.}
	\label{fig:conf_dens_pot}
\end{figure}

\begin{figure}
	\centering
	\begin{tabular}{ccc}	
		\begin{overpic}[width=0.32\columnwidth]{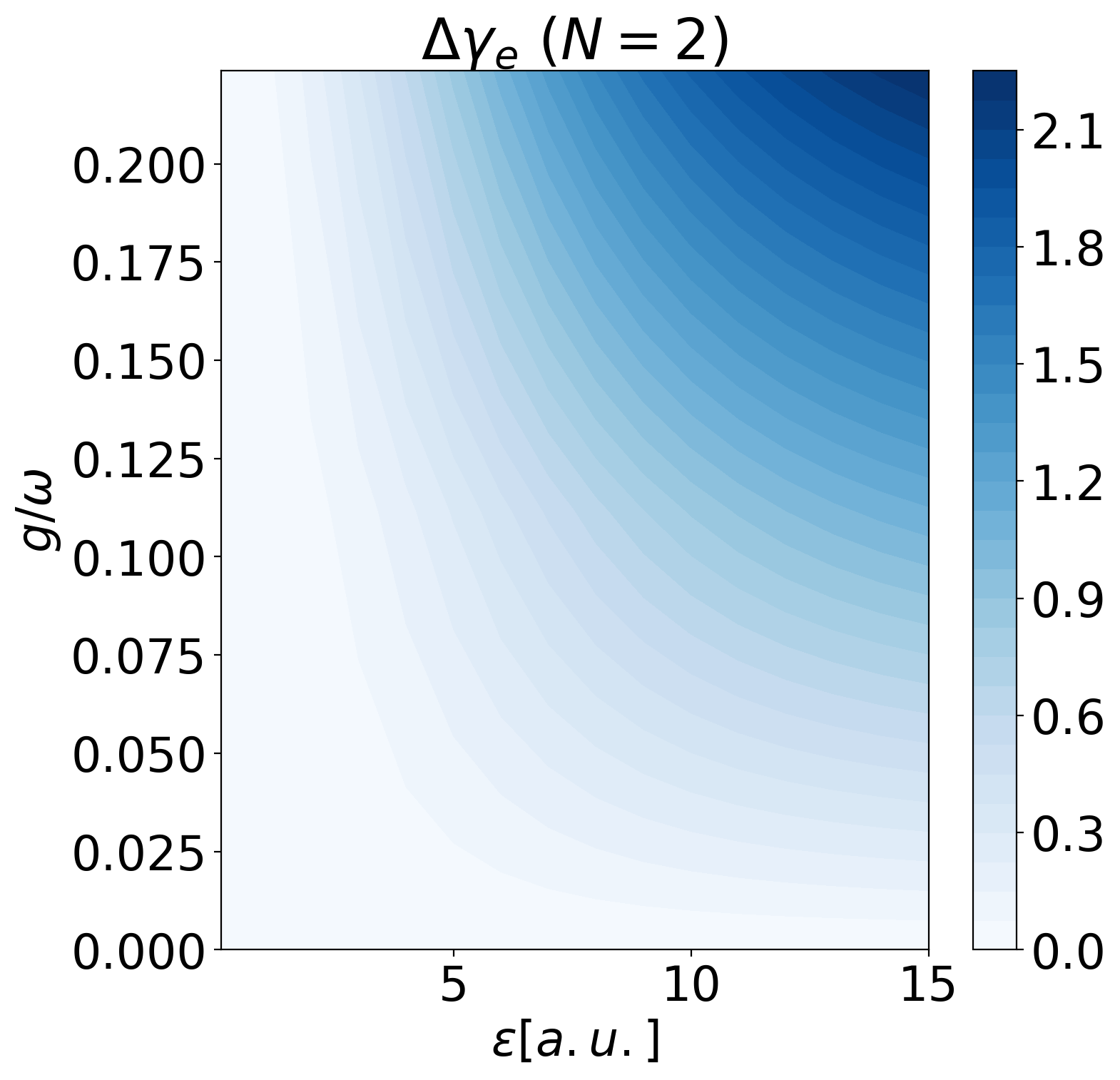}
			\put (21,80) {\textcolor{black}{(a)}}
		\end{overpic}
		&
		\begin{overpic}[width=0.32\columnwidth]{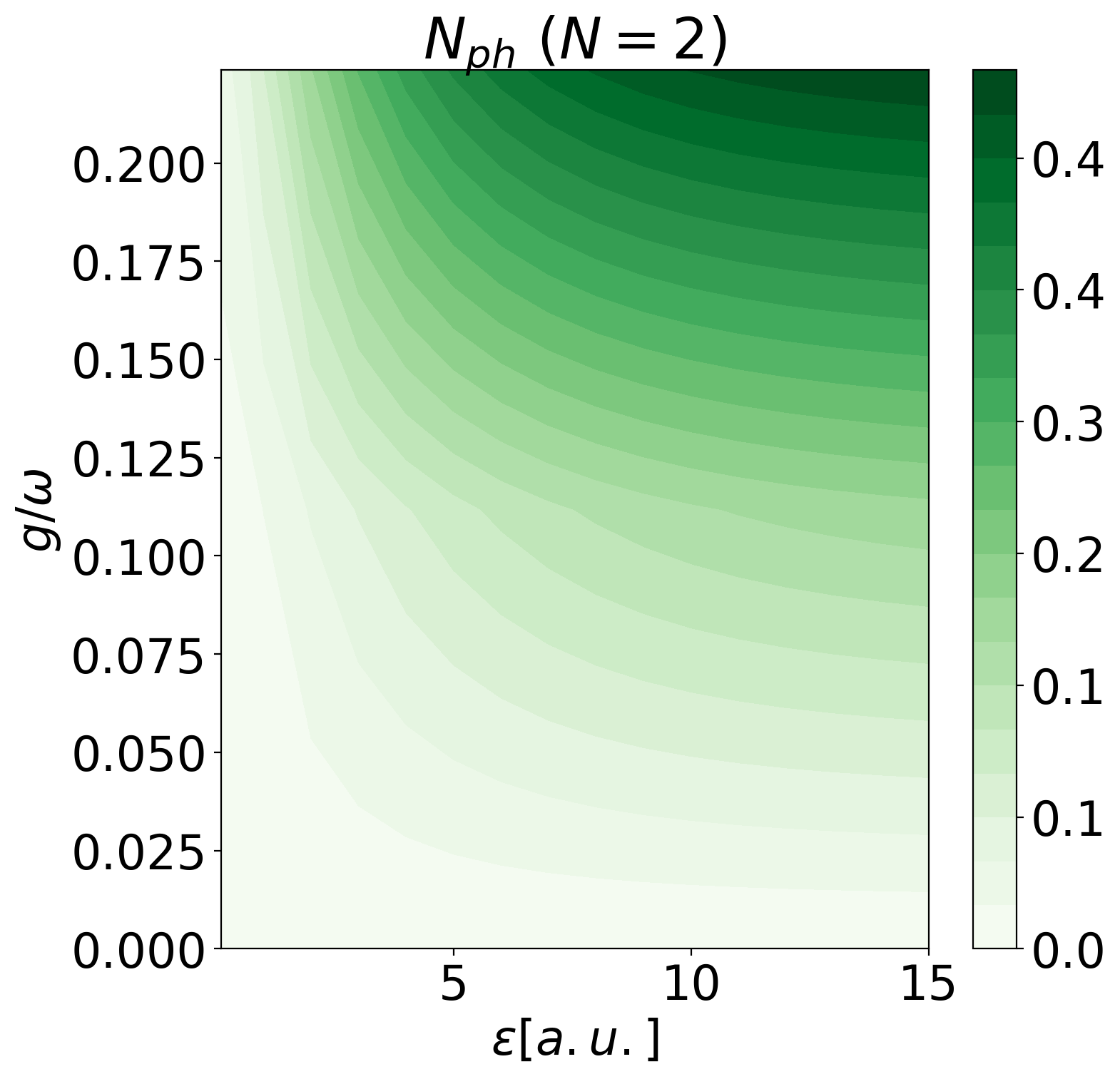}
			\put (21,80) {\textcolor{black}{(b)}}
		\end{overpic}
		&
		\begin{overpic}[width=0.32\columnwidth]{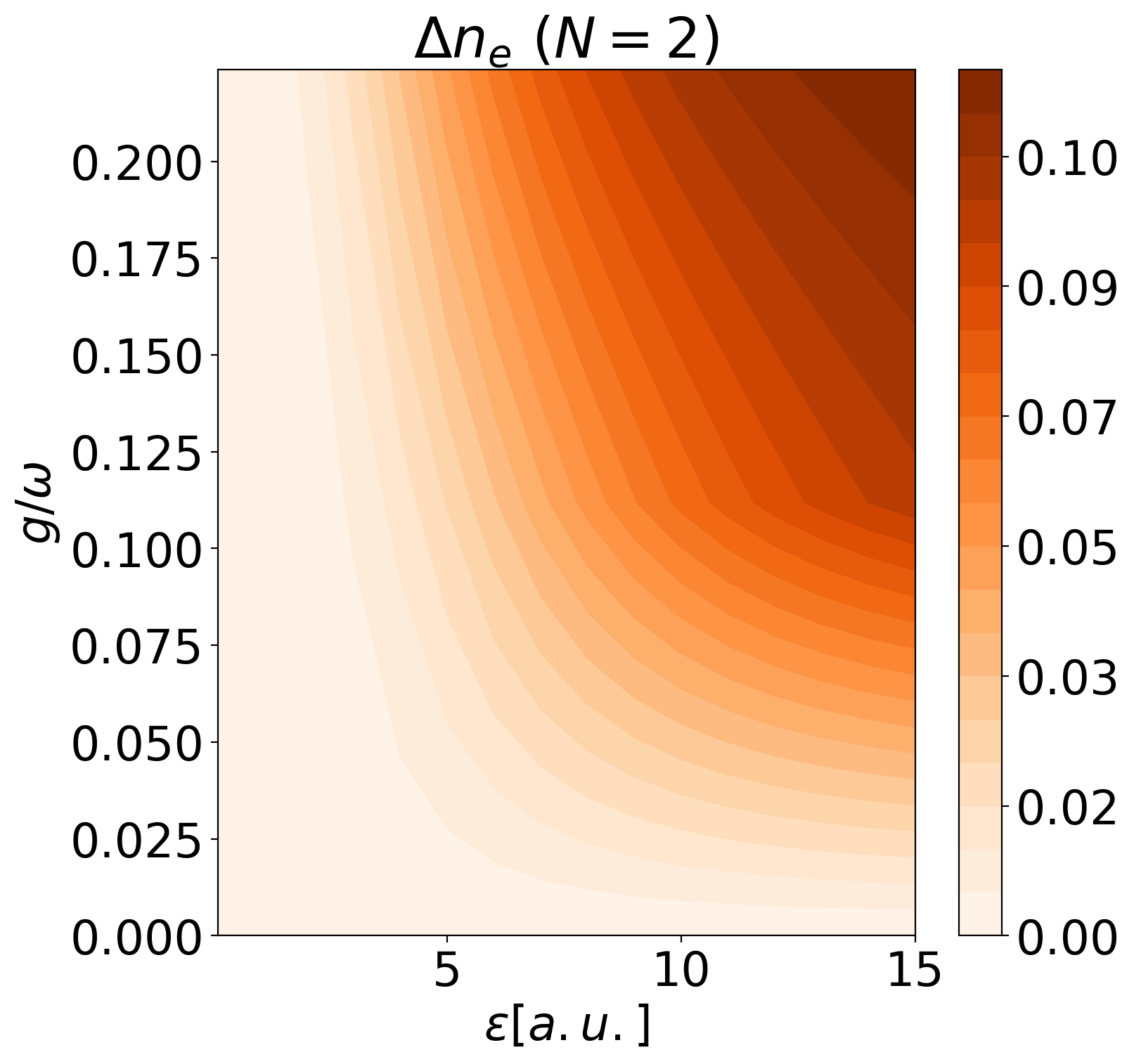}
			\put (21,78) {\textcolor{black}{(c)}}
		\end{overpic}
		\\
		\begin{overpic}[width=0.32\columnwidth]{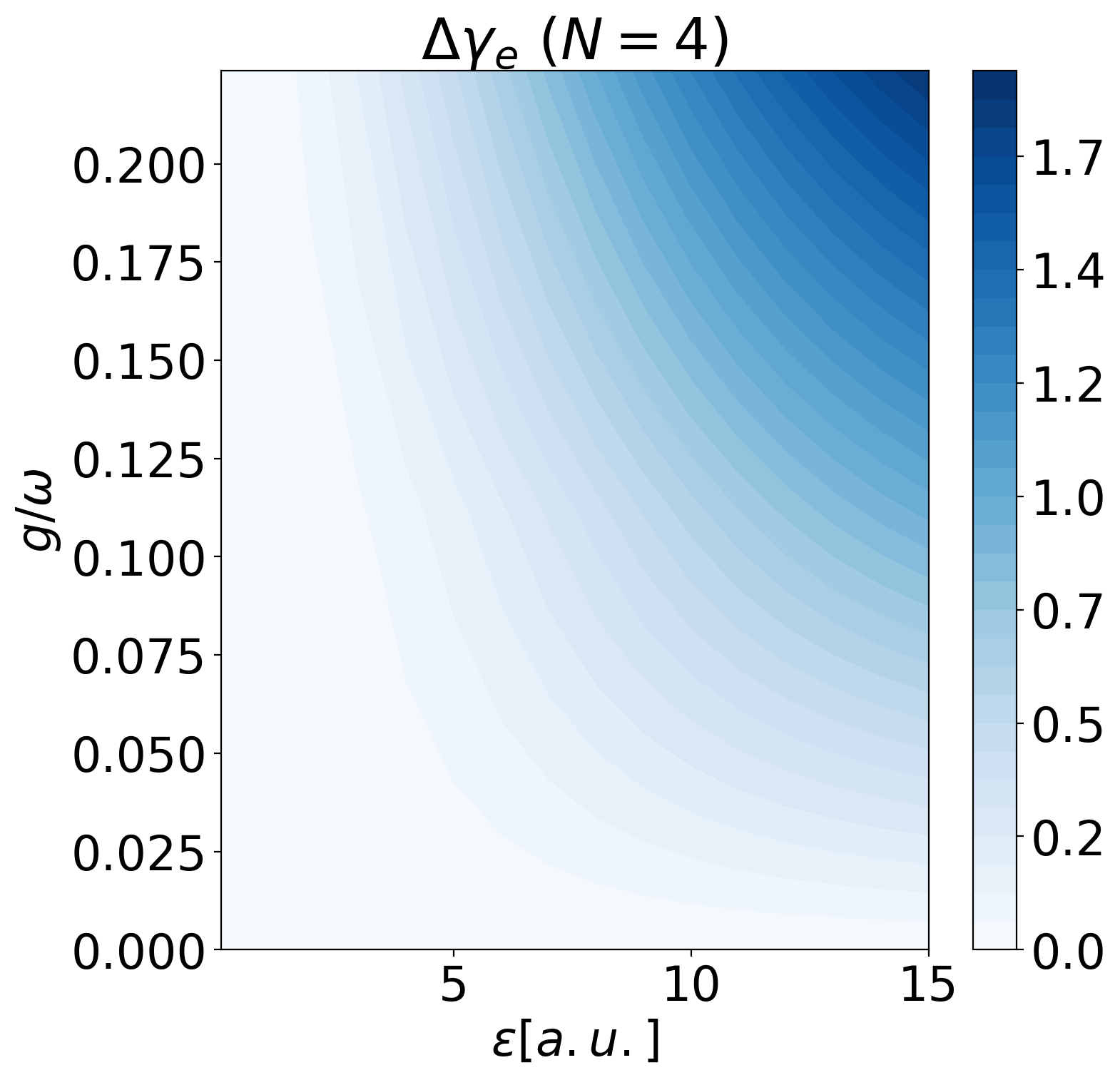}
			\put (21,80) {\textcolor{black}{(d)}}
		\end{overpic}
		&
		\begin{overpic}[width=0.32\columnwidth]{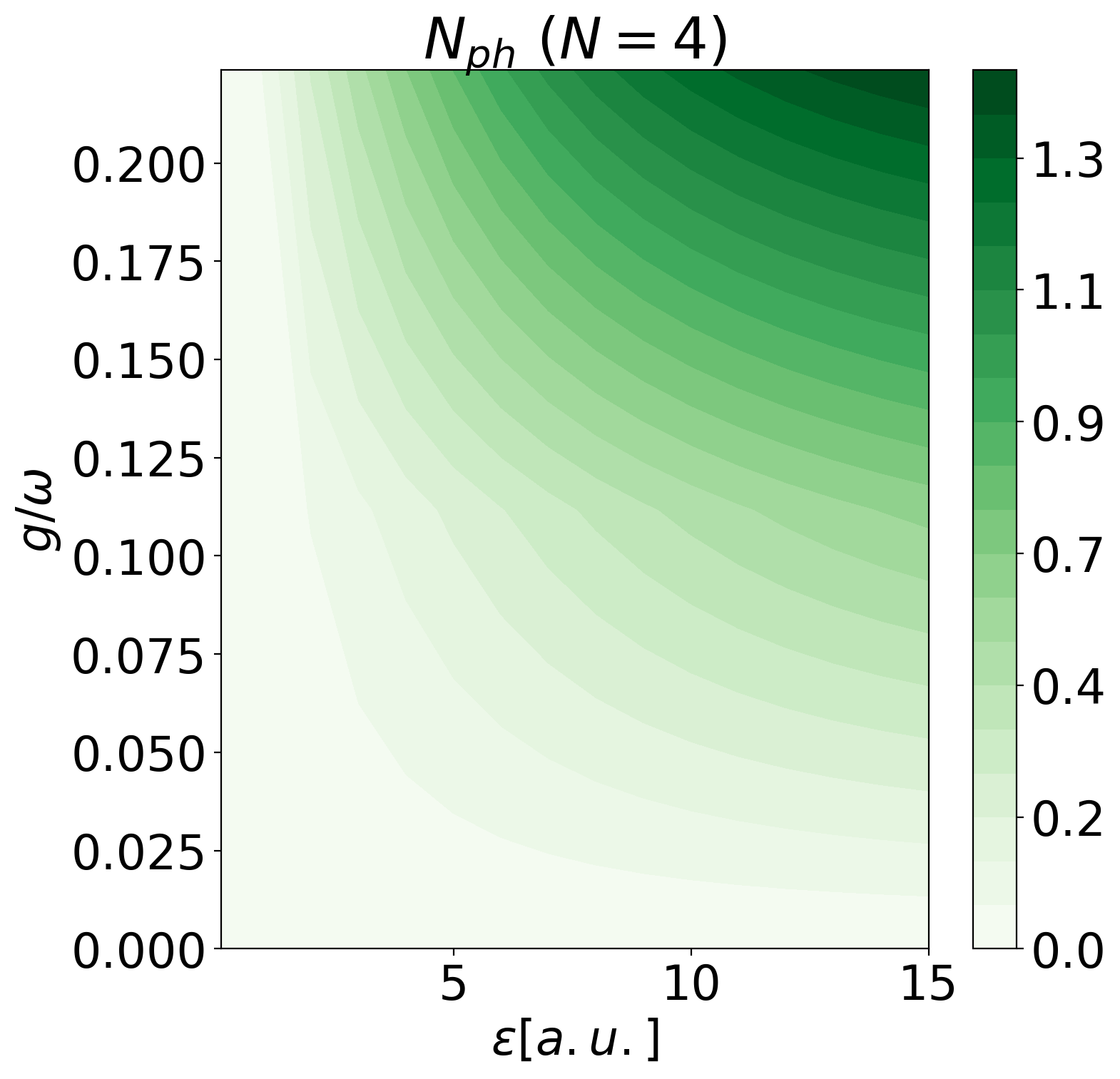}
			\put (21,80) {\textcolor{black}{(e)}}
		\end{overpic}
		&
		\begin{overpic}[width=0.32\columnwidth]{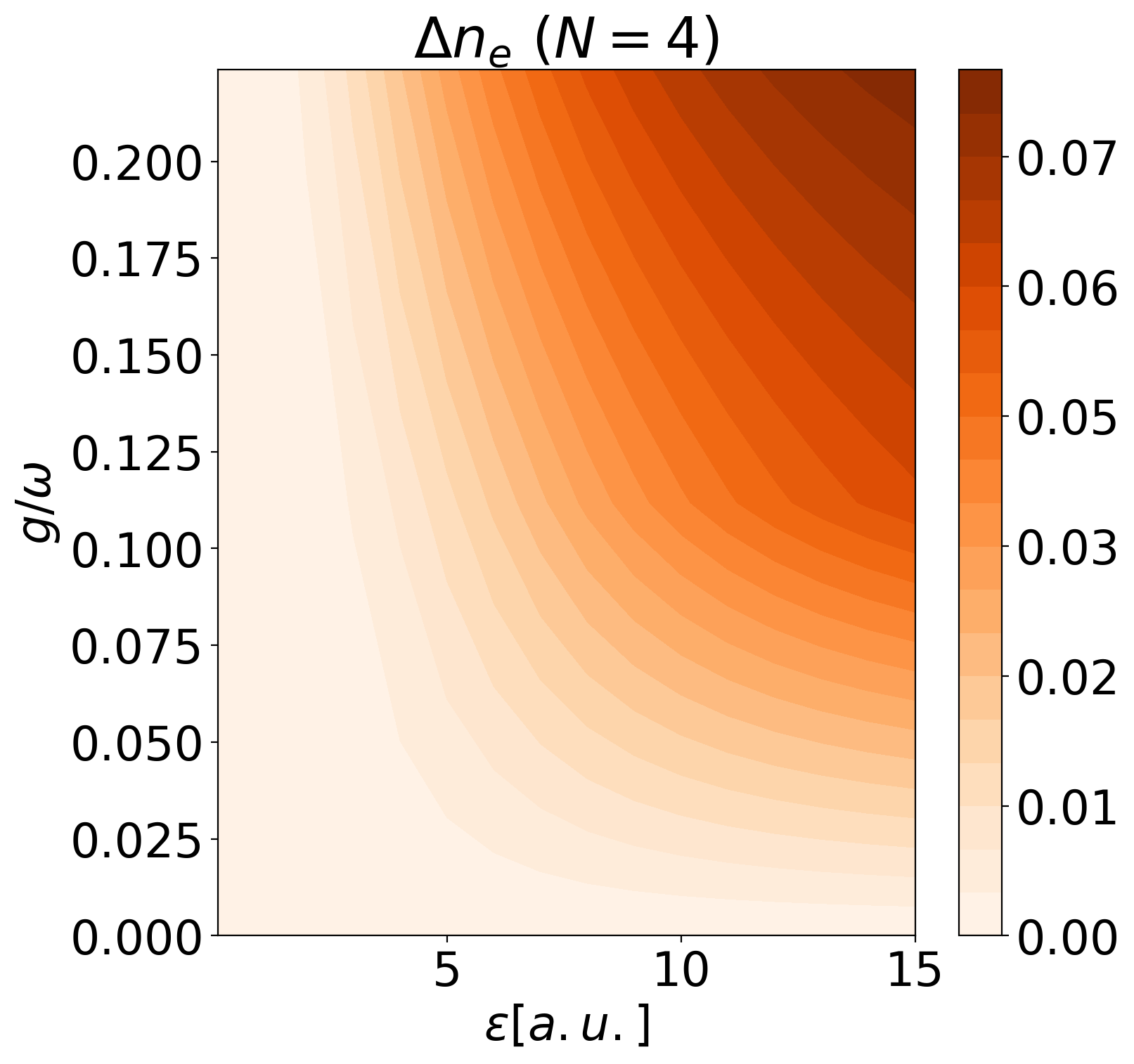}
			\put (21,78) {\textcolor{black}{(f)}}
		\end{overpic}
	\end{tabular}
	\caption{The plots show key quantities of the model system as a function of the coupling strength $g/\omega$ and the localization parameter $\epsilon$ for the 2-electron (upper row) and 4-electron (lower row) case. In the first column, we depict the normalized deviation $\Delta \gamma_e=||\gamma^e_{g/\omega,\epsilon} - \gamma^e_{g/\omega=0.0,\epsilon}||_2/N$ of the electronic 1RDM  to a reference for the same $\epsilon$ and $g/\omega=0$ (blue). In the second column, we show the total photon number $N_{ph}$ (green) and in the third column, the total deviation of the electronic natural occupation numbers $\Delta n_e =\sum_i ||n^e_{i, g/\omega,\epsilon} - n^e_{i, g/\omega=0.0,\epsilon}||_2$ is displayed. This is a measure of the induced electron-photon correlation. We observe in all the cases that when the bare matter wave function becomes more delocalized ($\epsilon$ larger), the modifications due to the matter-photon coupling become stronger.}
	\label{fig:conf_contour}
\end{figure}

\begin{figure}[h]
	\centering
	\includegraphics[width=0.49\columnwidth]{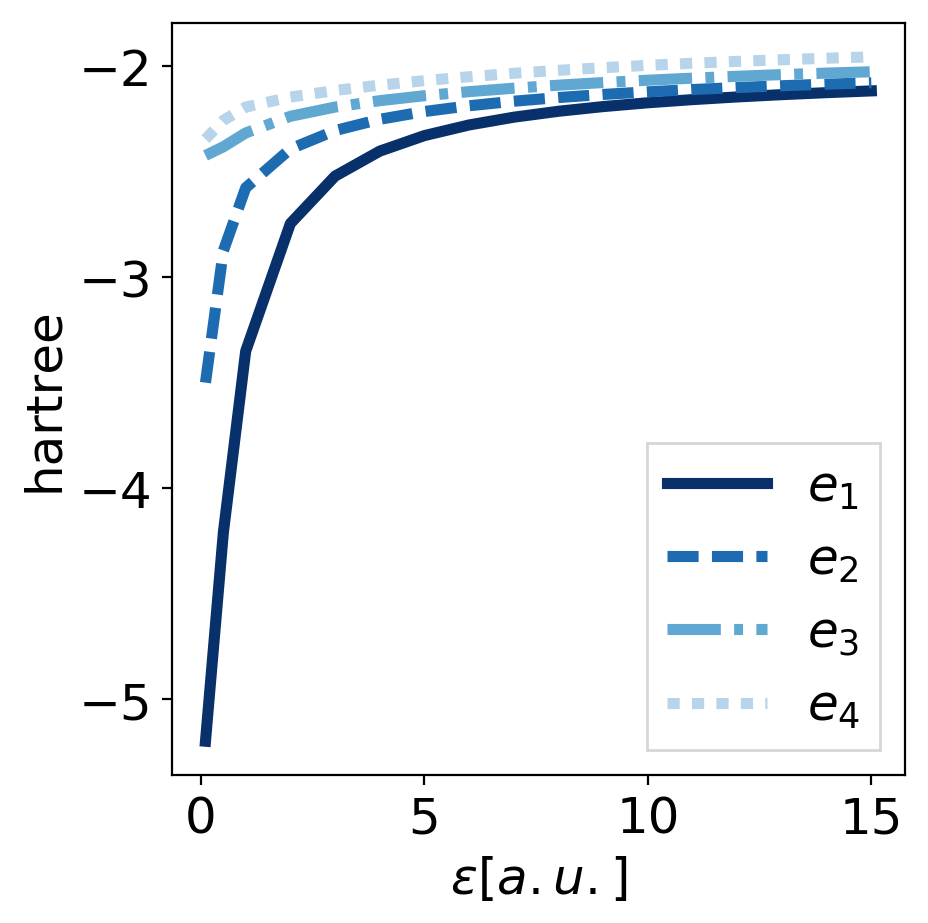}
	
	\caption{The first four eigenenergies $e_1, \dots ,e_4$ of the electronic one-body equation $[-\frac{1}{2}\partial_x^2- v(r)]\psi_i=e_i\psi_i$ for $N=2$ are shown as a function of the confinement parameter $\epsilon$. We see that the energies approach each other with increasing $\epsilon$ and thus decreasing confinement. This means that for a fixed coupling strength, the less confined the system is the more states are available in a small energy range for photon-induced modifications of the ground state.}
	\label{fig:electronic_energy}
\end{figure}

Having illustrated the need to enforce the correct symmetry for the use of the polaritonic basis, we employ polaritonic HF to study the effect of electron-photon coupling versus electron localization. We consider a matter system with a local potential $v(x)=N/\sqrt{x^2 + \epsilon^2}$, which represents a potential well that is deep (shallow) for small (large) $\epsilon$. The parameter $\epsilon$ thus represents the level of confinement of the potenial $v(x)$ which is depicted in green (for various values of $\epsilon$) in Fig.~\ref{fig:conf_dens_pot}.   
To reduce boundary effects (which also represent a form of confinement), we consider a box length of 30 bohr ($B_m=30$), corresponding to a real-space grid from $x=-14.5$ bohr to $x=14.5$ bohr. We consider the case of a 2-electron and a 4-electron system, respectively, and set $\omega=0.1$ hartree, which is far away from the regime where the fermionic HF approximation is valid. Thus the right hybrid statistics of the polaritons are crucial. We consider again $B_{ph}=5$, for which all the results are converged. We want to stress here that the corresponding many-body space for 4 particles has a dimension of $(2B_{s})^{N}*B_{ph}= 64.8\cdot10^6$ and thus is practically \emph{inaccessible} by exact diagonalization.
Let us first consider how the electronic ground-state density changes when coupling and the localization are varied. To facilitate the comparison between the $N=2$ and $N=4$ case we plot in Fig.~\ref{fig:conf_dens_pot} the normalized electronic ground-state density $\rho(x)/N$, where $\rho=\gamma_e(x,x)$ is the diagonal of the electronic 1RDM (in blue) and the normalized confinement $v(x)/N$ (in green). In Fig.~\ref{fig:conf_dens_pot} (a) we show the uncoupled 2-particle case and in (b) we use $g/\omega = 0.2$ for the 2-particle case for varying $\epsilon$. In Fig.~\ref{fig:conf_dens_pot} (c) and (d) we show the same plots for the 4-particle case. In both cases we see that for strongly-confined electrons, i.e.\ for small values of  $\epsilon$, the influence of the strong light-matter coupling on the density is negligible. This is in agreement with the usual assumption underlying, e.g.\ the Jaynes--Cummings model, that the ground state for atomic systems is only slightly affected by coupling to the photons of a cavity mode. Much higher coupling strengths would need to be employed in order to see a sizeable effect for strong localization. In contrast, once we lift the confinement and the electrons get delocalized, the influence of the light-matter coupling becomes appreciable. The induced changes are not uniform but depend on the details of the electronic structure, i.e.\ in the $N=2$ case we have a clear localization effect (Fig.~\ref{fig:conf_dens_pot} (b)) while for $N=4$ we have an enhancement or even emergence of the double-peak structure (Fig.~\ref{fig:conf_dens_pot} (d)). That the changes in the electronic structure due to strong electron-photon coupling are non-trivial are in agreement with previous studies~\cite{Flick2018a,Buchholz2019}. 
To make these observations more quantitative we display in Fig.~\ref{fig:conf_contour} the normalized changes in the electronic 1RDMs depending on the confinement and the coupling strength, i.e.\ $\Delta \gamma_e = \lVert \gamma^e_{g/\omega, \epsilon} - \gamma^e_{g/\omega =0, \epsilon}\rVert_2 /N$, in panel (a) and (d) for the 2-particle case and the 4-particle case, respectively. Also, we show the photon number $N_{ph}$ in the ground state in dependence of confinement and coupling strength in (b) and (e) for the 2-particle case and the 4-particle case, respectively. As a third quantity we consider $\Delta n_e =\sum_i ||n^e_{i, g/\omega,\epsilon} - n^e_{i, g/\omega=0.0,\epsilon}||_2$, where $n_i$ are the natural occupation numbers. For the zero-coupling case they are all either zero or one, which corresponds to a single Slater determinant in the electronic subspace. If they are between zero and one they indicate a correlated (multi-determinantal) electronic state. Therefore $\Delta n_e$ measures the photon-induced correlations and also highlights that although polaritonic HF is a single-determinant method in the polaritonic space, for the electronic system it is a correlated (multi-determinantal) method. For both, the 2- and the 4-particle case we find consistently that the more delocalized the uncoupled matter system is, the stronger the coupling modifies the ground state. Although this effect depends on the details of the electronic structure as we saw in Fig.~\ref{fig:conf_dens_pot}, the plots of Fig.~\ref{fig:conf_contour} indicate that this behavior is quite generic. The reason is that within a small energy range many states with different electronic configurations are available as opposed to a strongly bound (and hence energetically separated) ground-state wave function (see Fig.~\ref{fig:electronic_energy}). A glance on the correlation measure $\Delta n_e$ in panel (c) and (f) of Fig.~\ref{fig:conf_contour} strengthens this explanation: For large $\epsilon$ and $g/\omega$ the electronic correlation is strongest and thus many electronic configurations contribute to the states of this parameter regime. This indicates also that the effective one-body description of the ground state of many cavity-QED models might be inaccurate in this regime.
Additionally, we observe that the maximal values of $\Delta \gamma_e$ in the 2-particle case are slightly larger than in the 4-particle case, which again is related to the different electronic structures of the two systems. 

For the photon numbers, however, which are depicted in panel (b) and (d) of Fig.~\ref{fig:conf_contour}, we see that in general the number of photons is larger in the 4-particle case. This is due to the simple reason that the more charge we have the more photons are created. Nevertheless, the amount of photons does not just double (as expected from a simple linear relation) but is almost three times higher. This highlights the non-linear regime of electron-photon coupling that we consider here. Again the number of photons increase also with the delocalization and hence the parameter $\epsilon$ is a very decisive quantity.
All these results point towards an interesting parameter in the context of strong light-matter coupling: the localization of the matter wave function. In agreement with a recent case study for simple 2-particle problems~\cite{Schafer2019}, systems that are less confined react much stronger to a cavity mode. This does not only hold for the ground state~\cite{Schafer2019}, but suggests that if we want to observe genuine modifications of the ground state due to strong light-matter coupling, we should consider matter systems that have a spatially extended wave function.  
One way would be large molecular or solid-state systems, the other way would be an ensemble of emitters. In the latter case the strong influence would lead to local changes, in contrast to the Dicke-like description of collective strong coupling, where we have $N$ independent replicas of the same, perfectly localized system. In both cases, it seems plausible that there are strong modifications of the electronic structure even if there is only coupling to the vacuum of a cavity. Modifications of chemistry by merely the vacuum seem therefore to be feasible. At the same time an interesting perspective with respect to collective strong-coupling arises: Maybe it is not the simple Dicke-type collectivity\cite{Kirton2019}, where an excitation is delocalized over many replica, that drives the changes in chemistry, but rather a genuine cavity-photon mediated spatial delocalization of the ensemble wave function. To answer this question, further investigations especially for realistic three-dimensional systems and ensembles including the Coulomb interaction have to be conducted. And polaritonic first-principle methods as introduced in this work seem specifically well-suited to answer these interesting and fundamental questions.

\section{Conclusion and Outlook}
\label{sec:conclusion}

In this article, we have highlighted the influence of the hybrid Fermi-Bose statistics of the polariton wave function in the dressed construction. This dressed construction allows to use (matter-only) first-principle methods to investigate strong light-matter coupled systems. We have provided a simple and general prescription on how to turn a given electronic-structure method into a polaritonic-structure method by introducing conditions in terms of the 1RDM to ensure the right hybrid statistics. We have given several examples where ignoring the correct Fermi-Bose statistics leads to unphysical results. As a specific example of a polaritonic-structure method we have transformed electronic HF theory to polaritonic HF theory and have demonstrated numerically that this approach stays accurate even for very strong coupling between light and matter. Finally, we have applied polaritonic HF to investigate the influence of electron localization in the context of strong electron-photon coupling and found for non-trivial examples that the more delocalized the uncoupled matter wave function is, the stronger it reacts to the modes of a cavity, which comes along with an increase of electronic correlation. This result indicates that there might be strong modifications of the ground state for spatially extended systems even by only coupling to the vacuum of a cavity. Specifically this raises the question whether an ensemble of emitters with a spatially extended wave function might show collective strong-coupling effects that also modify the local electronic structure, in contrast to the Dicke-type collective coupling picture that does not consider electronic correlation.

The presented results provide a relatively straightforward way to perform simulations for strong light-matter coupling based on well-established theoretical and numerical methods. The main two practical bottlenecks are the increase of the dimension of the basic orbitals of the theory and the inclusion of the new constraints that enforce the right statistics. The former bottleneck will in most cases mean that one has instead of 3-dimensional orbitals now 4-dimensional ones, since it is usually a single effective mode that is considered, and is therefore not overly numerically expensive. It is crucial to realize here that, besides the necessarily larger orbital bases, the scaling of a method is not effected by the inclusion of the photon field~\footnote{Note that we do not want to underestimate the influence of an entirely new dimension in the problem. The given statement depends of course on the necessary size of the photon-basis. However, in the cases that we studied so far, we observed the converged results with photon bases that where of the order of the number of particles.}. The second bottleneck is instead challenging algorithmic-wise. As implemented for the test examples, one has to construct the full dressed 1RDM and then determine the electronic 1RDM as well as diagonalize it in each iteration step of the self-consistency cycle. Further, for each orbital one needs an extra Lagrange multiplier in the augmented-Lagrangian method. These two things taken together as well as the non-linearity of the constraint lead to a quite slow convergence. Yet, there are different methods to enforce the new constraints and also the existing implementation can be made much more efficient. To extend, for example, the existing implementation of the dressed construction in the real-space electronic-structure code \textsc{Octopus}~\cite{Tancogne-Dejean2020} to enforce the hybrid statistics, one would not explicitly diagonalize the electronic 1RDM, which is numerically prohibitively expensive but resort to \emph{semi-definite programming} methods~\cite{Vandenberghe1996} that exploit the reformulation of the constraints in terms of positivity-conditions (see Sec.~\ref{sec:est}).\footnote{Note that independently of the precise algorithm, the bottleneck in a full real-space description of polaritonic orbitals is the extra dimension. However, with modern high-performance clusters, systems that are not too large should be manageable even in 4D.}. In orbital-based codes, our in Sec.~\ref{sec:est} presented algorithm could be well applicable, if not too large systems are considered and a direct diagonalization of the electronic 1RDM is feasible. Note that in contrast to the matter-description, the extra photon orbitals do not necessarily have to be constructed in the code. The analytically known structure of the photon-number states allows to calculate all necessary matrix elements analytically. We have exploited this also in our implementation (see Sec.~\ref{sec:study_case}). 

We believe that such extensions, though they require a certain amount of work, are indeed worthwhile. The presented results for a simple yet still numerically challenging model system highlight that much is still to be understood and discovered in the context of polaritonic chemistry and physics. Despite the great success that cavity QED models have in explaining certain aspects of this emerging field, similar to condensed-matter physics, genuine first-principle methods seem to be necessary to get a detailed understanding. Many of the exciting experimental results, like the changes in chemical reactions, still are poorly understood and the basic mechanisms are yet to be discovered. To find explanations that are as unbiased as possible, as well as to make quantitative predictions calls for the availability of genuine first-principle methods for coupled matter-photon systems. At this point, we want to stress that we considered polaritonic HF just as a test-case to get a basic understanding of the hybrid statistics from a physical but especially from a numerical perspective. This understanding is the basis to implement the whole machinery in a standard electronic-structure code like \textsc{Octopus}. Once, this is accomplished, we will be able to perform polaritonic RDMFT and QEDFT calculations. Previous works~\cite{Nielsen2018,Buchholz2019} showed that both levels of theory seem very promising for an accurate description of the strong coupling between molecular systems and a cavity mode. The main open gap of these methods was a numerically feasible inclusion of the hybrid statistics, which we closed with this work.

Let us finally also comment on possible applications of the above construction for other multi-species situations. It seems possible, provided we can define species with the same number of particles, that one can instead of working with complicated multi-species wave functions, work with the combined density matrices and enforce the ensemble representability conditions on the subsystems. This does not necessarily need any dressed construction. For instance, think about the Schr\"odinger equation for electrons and nuclei/ions. Assuming that we have one kind of nuclei/ions we could express the combined density matrix in terms of electron-nuclei/ion pairs. It seems interesting to investigate the above procedure also in the context of such cases.  

\begin{acknowledgement}
	F.B. would like to thank Enrico Ronca and Christian Schäfer for stimulating and useful discussions. Additionally, F.B. would like to thank Wilhelm Bender for sharing his experience about nonlinear programming. This work was supported by the European Research Council (ERC-2015-AdG694097), the Cluster of Excellence "Advanced Imaging of Matter"(AIM), Grupos Consolidados (IT1249-19) and SFB925. The Flatiron Institute is a division of the Simons Foundation.
\end{acknowledgement}

\appendix
\section{The exact mapping between the physical and auxiliary wave function for two electrons and one mode}
\label{app:mapping}
The ground state of Hamiltonian \eqref{eq:Hamiltonian} for two electrons and one photon mode is of the form $\Psi(\bx_1,\bx_2,p)=\sum_{i,j,\alpha}C_{ij}^{\alpha}\psi_i(\bx_1)\psi_j(\bx_2)\chi^{\alpha}(p)$ for a given electronic spin-spatial one-particle basis $\{\psi_i(\bx)\}\, (\bx=(\br \sigma)) $ and a photonic basis $\{\chi^{\alpha}(p)\}$, that we specifically choose as eigenfunctions of the photon Hamiltonian $\hat{H}_{ph}$, defined in Sec. \ref{sec:physical_setting}. We map $\Psi$ to the auxiliary ground state
\begin{align*}
\Psi'(\bx_1,\bx_2,p,p_2)=\left[\sum_{i,j,\alpha}C_{ij}^{\alpha}\psi_i(\bx_1)\psi_j(\bx_2)\chi^{\alpha}(p)\right]\chi^0(p_2),
\end{align*}
where $\chi^0(p_2)$ is the vacuum state of the auxiliary harmonic oscillator, which by construction has the same frequency and thus is the lowest eigenstate of $\hat{H}_{ph}$. Then, we perform an orthogonal coordinate transformation, c.f. Eq.~\eqref{eq:CenterOfMass}, which for the two-particle case reads
\begin{align*}
\Psi'(\bx_1,\bx_2,p,p_2)\longrightarrow \Psi'(\bx_1,\bx_2,p=1/\sqrt{2}(q_1+q_2),p_2=1/\sqrt{2}(q_1-q_2)).
\end{align*}
Since we know the analytical expression of the photon basis, i.e. $\chi^{\alpha}(p)=\frac{1}{\sqrt{2^{\alpha}\alpha!}}\left(\frac{\omega}{\pi}\right)^{1/4} e^{-\omega p^2/2} H_{\alpha}(\sqrt{\omega}p)$, where $H_{\alpha}(z)=(-1)^{\alpha}e^{z^2}\frac{\td^{\alpha}}{\td z^{\alpha}}\left(e^{-z^2}\right)$ are Hermite-polynomials, we can perform this coordinate transformation explicitly. Specifically, we have to calculate terms of the form 
\begin{align*}
	\chi^{\alpha}(1/\sqrt{2}(q_1+q_2))\chi^0(1/\sqrt{2}(q_1-q_2)) \propto e^{-\omega (q_1+q_2)^2/4} H_{\alpha}(\sqrt{\omega/2}(q_1+q_2)) e^{-\omega (q_1-q_2)^2/4}
\end{align*}
for all $\alpha$, where we used already that $H_0=1$. This is trivial for the exponential part of the oscillator-states, since $e^{-\omega p^2/2}e^{-\omega p_2^2/2}\rightarrow e^{-\omega (q_1+q_2)^2/4}e^{-\omega (q_1-q_2)^2/4}=e^{-\omega q_1^2/2}e^{-\omega q_2^2/2}$. For the remaining part involving the Hermite-polynomial, we use the identity $H_{\alpha}(z_1+z_2)=2^{-\alpha/2}\sum_{\beta=0}^{\alpha} {\alpha\choose\beta}H_{\alpha-\beta}(z_1\sqrt{2})H_{\beta}(z_2\sqrt{2})$ for $z_i=\sqrt{\omega/2}q_i$. After some algebra we arrive at
\begin{align}
\label{eq:Psi_q}
\Psi'(\bx_1,\bx_2,q_1,q_2)=\sum_{i,j,\alpha}C_{ij}^{\alpha}\psi_i(\bx_1)\psi_j(\bx_2) \sum_{\beta=0}^{\alpha} \sqrt{\frac{\alpha!}{\beta!(\alpha-\beta)!}}\frac{1}{\sqrt{2^{\alpha}}} \chi^{(\alpha-\beta)}(q_1)\chi^{\beta}(q_2).
\end{align}
This is in a similar way possible for every number of modes and particles, but will involve considerably more cumbersome expressions. However, this easiest non-trivial examples exhibits already all the features of the auxiliary construction. For instance, we can highlight here, how the ``information'' of coordinate p is distributed uniformly over the new coordinates $q_1$ and $q_2$.

\section{The symmetry problem of the polariton approximation}
\label{app:symmetry}
In this appendix, we elucidate the problems of a straightforward application of the polariton symmetry of Eq. \eqref{eq:SymmetryPauliDressedFermion} on the level of the many-body basis, i.e. generalizing the concept of a Slater determinant to polaritonic \emph{and} electronic coordinates. For two particles, such a generalized determinant is given by
\begin{align*}
	\Psi_{ab}'(\bz_1,\sigma_1,\bz_2,\sigma_2)&= \phi_{a}(\br_1,q_1,\sigma_1)\phi_{b}(\br_2,q_2,\sigma_2)- \phi_{b}(\br_1,q_1,\sigma_1)\phi_{a}(\br_2,q_2,\sigma_2)\\
	&+ \phi_{a}(\br_1,q_2,\sigma_1)\phi_{b}(\br_2,q_1,\sigma_2)- \phi_{b}(\br_1,q_2,\sigma_1)\phi_{a}(\br_2,q_1,\sigma_2),
\end{align*}
where $\phi_{a/b}$ are some (orthonormal) polariton orbitals. Let us try to calculate the norm-square of $\Psi_{ab}'$, which reads
\begin{align*}
||\Psi_{ab}'||^2=& \sum_{\sigma_1,\sigma_2}\int\td\bz_1\bz_2 \Psi_{ab}'{}^*(\bz_1,\sigma_1,\bz_2,\sigma_2) \Psi_{ab}'(\bz_1,\sigma_1,\bz_2,\sigma_2)\\
=& 4 \braket{\phi_a|\phi_a}\braket{\phi_b|\phi_b} -4 \braket{\phi_a|\phi_b}\braket{\phi_b|\phi_a} \\
& +\sum_{\sigma_1,\sigma_2}\int\td\bz_1\bz_2 \phi_{a}^*(\br_1,q_1,\sigma_1)\phi_{b}^*(\br_2,q_2,\sigma_2) \phi_{a}(\br_1,q_2,\sigma_1)\phi_{b}(\br_2,q_1,\sigma_2) + c.c. \\
&- \sum_{\sigma_1,\sigma_2}\int\td\bz_1\bz_2 \phi_{a}^*(\br_1,q_1,\sigma_1)\phi_{b}^*(\br_2,q_2,\sigma_2) \phi_{b}(\br_1,q_2,\sigma_1)\phi_{a}(\br_2,q_1,\sigma_2) - c.c.\\
& + \sum_{\sigma_1,\sigma_2}\int\td\bz_1\bz_2 \phi_{b}^*(\br_1,q_1,\sigma_1)\phi_{a}^*(\br_2,q_2,\sigma_2) \phi_{b}(\br_1,q_2,\sigma_1)\phi_{a}(\br_2,q_1,\sigma_2) + c.c. \\
&- \sum_{\sigma_1,\sigma_2}\int\td\bz_1\bz_2 \phi_{b}^*(\br_1,q_1,\sigma_1)\phi_{a}^*(\br_2,q_2,\sigma_2) \phi_{a}(\br_1,q_2,\sigma_1)\phi_{b}(\br_2,q_1,\sigma_2) - c.c.
\end{align*}
In this expression, only the first line is what would appear in a standard Slater determinant, i.e. overlaps of orthonormal orbitals with a constant norm for all $\phi_{a/b}$.
However, with the terms that stem from the additional symmetry requirements, many new ``mixed-index'' terms arise, whenever one or more of the according orbitals have coordinates with different indices. All these 8 terms are two-body like integrals, which is in stark contrast to the calculation of the norm for a normal Slater determinant that involves only one-body terms. Their computation is non-trivial, which also is contrast the normal terms of the first line, which are analytically given by the orthonormalization condition. The occurrence of such terms also means that the norm of $\Psi_{ab}'$ depends on the specific form of $\phi_{a/b}$ and thus has be calculated explicitly to normalize $\Psi_{ab}'$.
The number of such terms for an N-body generalized determinant is given by the possible permutations of polaritonic and electronic coordinates $N!^2$ minus the $N!$ ``ordinary'' terms and grows factorial with the number of particles, i.e.\ $N!^2-N!$. We see that the normalization and in the same way also the calculation of expectation values of a wave function that explicitly exhibits the symmetry \eqref{eq:SymmetryPauliDressedFermion} requires the numerical calculation of (over)exponentially many non-trivial terms.
This explicit ansatz is thus infeasible in practice and instead one should use more efficient ways to enforce the polariton symmetry as discussed in section \ref{sec:dressed_orbitals}.

\section{Numerical Details}
\label{app:numerical_details}
For the numerical results of Sec. \ref{sec:study_case}, we wrote a Python code relying mainly on the routines of the package NumPy.\footnote{See \url{https://www.python.org/} and \url{https://numpy.org/} for further information.}
The code specifically constructs the one-body Hamiltonian, defined in Eq. \eqref{eq:PdHF_Fockmatrix} allowing in principle for an arbitrary lattice basis for the matter system (in this publication we always considered a one-dimensional real-space grid, but one could also consider, e.g.\ atomic-orbital and implement the respective matrix elements) and Fock-number states for the one photon-mode. The minimization routine of the code is based on the conjugate-gradient algorithm described in \citet{Payne1992}, Chap. V, replacing the Lagrangian and the gradient expression by \eqref{eq:Lagrangian} and \eqref{eq:GradientPhi}, respectively. The extra parameters and inner loop convergence criteria due to the new constraints \eqref{eq:ConstraintsInequality} are updated according to Algorithm 17.4, p. 520 of \citet{Nocedal2006}. To include the second gradient, c.f. \eqref{eq:GradientPsiGamma}, we introduced a further loop, in which the electronic 1RDM is diagonalized. We used for all calculations the overall convergence criterium of max$(||\eqref{eq:GradientPhi}||, |E^{m}-E^{m-1}|) < 10^{-4}$, where $E^m$ is the total energy of the $m$-th iteration of the outer loop.

\bibliography{bibliography}


\end{document}